\documentclass[10pt,epsf]{article}

\usepackage{scicite}

\input psfig.sty

\def\gsim{\mathrel{\rlap{\lower 4pt \hbox{\hskip 1pt $\sim$}}\raise 1pt
\hbox {$>$}}}
\def\lsim{\mathrel{\rlap{\lower 4pt \hbox{\hskip 1pt $\sim$}}\raise 1pt
\hbox {$<$}}}

\newenvironment{sciabstract}{%
\begin{quote} \bf}
{\end{quote}}

\newcounter{lastnote}

\title{Protostellar Feedback Halts the Growth \\
of the First Stars in the Universe}

\author{
Takashi Hosokawa$^{1,2}$,  
Kazuyuki Omukai$^{2}$, 
Naoki Yoshida$^{3}$, and 
Harold W. Yorke$^{1}$ 
\\
\normalsize{$^{1}$Jet Propulsion Laboratory, California Institute
of Technology, Pasadena CA 91109, USA}\\ 
\normalsize{$^{2}$Department of Physics, Kyoto University,
Kyoto 606-8502, Japan}\\
\normalsize{$^{3}$Institute for the Physics and Mathematics 
of the Universe, Todai Institutes for Advanced Study,} \\
\normalsize{University of Tokyo, Kashiwa, Chiba
277-8568, Japan}\\
\\
\normalsize{$^\ast$To whom correspondence should be addressed; E-mail:
hosokwtk@gmail.com} \\
}

\date{}

\newcommand{\vvec}{\mbox{\boldmath$v$}}
\newcommand{\nvec}{\mbox{\boldmath$n$}}
\newcommand{\Fvec}{\mbox{\boldmath$F$}}
\newcommand{\Kvec}{\mbox{\boldmath$K$}}
\newcommand{\FFvec}{\mbox{\boldmath${\cal F}$}}

\begin{document} 



\maketitle 


\begin{sciabstract}
The first stars fundamentally transformed the early universe
by emitting the first light and by producing the first
heavy elements. These effects were predetermined by the mass
distribution of the first stars, which is thought to have been 
fixed by a complex interplay of gas accretion and 
protostellar radiation.
We performed radiation-hydrodynamics simulations that followed the 
growth of a primordial protostar 
through to the early stages as a star with thermo-nuclear burning. 
The circumstellar accretion disk was evaporated by ultraviolet  
radiation from the star when its mass was 43 times that
of the Sun. Such massive primordial stars, in contrast 
to the often postulated extremely massive stars, 
may help explain the fact that there are no signatures of the
pair-instability supernovae in abundance patterns of metal-poor
stars in our galaxy.
\end{sciabstract}


Theoretical studies and detailed computer simulations
show that the cradles of the first stars were dense concentrations 
of primordial gas, with masses of $\sim 1000$ that of the sun. 
Such gas clouds formed through radiative cooling, with hydrogen molecules 
at the center of a dark matter halo of $10^{6}$ solar mass ($M_\odot$), 
when the age of the universe was a few hundred
million years old\cite{BYHM09}.


According to our current understanding of star formation,
a gas cloud's dense core gravitationally contracts 
in a non-homologous run-away fashion, in which 
the densest parts become denser faster
than does the rest of the cloud\cite{ON98,ABN02,Y06,ON07}. 
In a primordial gas cloud, one or a few embryo
protostars are formed near the center\cite{ON98,Y08}.
The initial mass of these embryo protostars is only 
$\simeq 0.01~M_\odot$; the bulk of the  
dense core material remains in the surrounding envelope and is 
subsequently drawn toward the protostar (or protostars) through gravity. 
With the typical angular momentum of dense cores,  
the centrifugal barrier allows only a small amount of
infalling gas to accrete directly onto the star.
Instead, a circumstellar disk is formed and the gas is accreted
onto the central star through the disk\cite{YB99}.
The final mass of these first stars is fixed, when 
the mass accretion terminates.
However, when and how this termination occurs are largely unknown.


Because the luminosity increases rapidly with protostellar mass,
radiative feedback is expected to regulate 
the mass accretion and ultimately shut off the accretion flow, 
setting the final mass of the first stars.
A primordial star more massive than a few tens times
the mass of the Sun radiates  
a copious amount of hydrogen ionizing photons ($>13.6$~eV). 
As an accreting star grows in mass, the ionized region in its vicinity
grows, and eventually the circumstellar disk is directly 
exposed to the stellar ionizing radiation.  
The gas on the disk surface is photoionized and heated 
and evaporates away from the star-disk system.
A semi-analytical model of this process predicts that this 
photoevaporation quenches the accretion flow of the 
disk material and puts an end to the stellar growth\cite{MT08}. 
However, the interplay between the accretion flow and the stellar 
radiation is highly dynamical and complex.


To identify the exact mechanism that halts the growth of a first star 
and to determine its final mass, 
we applied a method used for studying the present-day 
massive star formation\cite{YW96,YS02}
to the case of the formation
of the first stars in a proper cosmological context.
We followed the radiation hydrodynamic evolution in the vicinity
of an accreting protostar, incorporating 
thermal and chemical processes in the primordial gas
in a direct manner.  
We also followed the evolution of the central protostar
self-consistently by solving the detailed structure 
of the stellar interior with zero metallicity as well as 
the accretion flow near the stellar surface
[supporting online material (SOM) text]\cite{OP03,HO09}. 
We configured the initial conditions by using the results
of a three-dimensional (3D)
cosmological simulation, which followed the entire history 
from primeval density fluctuations to the birth of 
a small seed protostar at the cosmological redshift 14\cite{Y08}.
Specifically, when the maximum particle number density 
reached $10^6~{\rm cm^{-3}}$ in the cosmological simulation, 
we considered a gravitationally bound sphere of radius 0.3~pc 
around the density peak, which enclosed a total gas mass of 
$\simeq 300~M_{\odot}$. We reduced the 3D data to an axisymmetric 
structure by averaging over azimuthal angles.                                   


The system was evolved until the central particle number 
density reached $10^{12}~{\rm cm^{-3}}$. We then  
introduced a sink cell of size $\simeq 10$ astronomical units (AU) 
and followed the subsequent evolution of the central protostar 
with a stellar evolution code.
The accretion rate onto the protostar was calculated directly
from the mass influx through the sink-cell boundary, whereas 
the luminosity from the protostar, which controls the 
radiative feedback, was calculated consistently from 
the protostellar model by using the derived accretion rate.


At its birth, a very small protostar of $\sim 0.01~M_{\odot}$ 
was surrounded by a molecular gas envelope of $\sim 1~M_\odot$, 
which was quickly accreted onto the protostar.  
Atomic gas further out initially had too much angular 
momentum to be accreted directly, and a circumstellar disk formed.
The infalling atomic gas first hit the disk plane roughly vertically 
at supersonic velocities.  A shock front formed; behind the shock,
the gas cooled and settled onto the disk, and its hydrogen
was converted to the molecular form via rapid gas phase
three-body reactions.
The molecular disk extended out to $\simeq 400$~AU from the 
protostar, when the stellar mass was $10~M_\odot$.
Accretion onto the protostar proceeded through this molecular disk
as angular momentum was transported outward.
The accretion rate onto the protostar was 
$\simeq 1.6 \times 10^{-3}~M_\odot~{\rm year}^{-1}$ at that moment.


The evolution of the central protostar is determined 
by competition between mass growth by accretion 
and radiative energy loss from the stellar interior. 
The time scale for the former is the accretion timescale
$t_{\rm acc} \equiv M_*/\dot{M}$, where $M_*$ is mass of the protostar
and $\dot{M}$ is mass accretion rate,
whereas that for the latter is 
the Kelvin-Helmholtz (KH) timescale 
$t_{\rm KH} \equiv G M_*^2/R_*L_*$, where $L_*$ 
is the luminosity from the stellar interior, $R_*$ is radius of the
star, and $G$ is gravitational constant.
The total luminosity of the protostar $L_{\rm tot}$ is the sum
of the stellar luminosity $L_*$ and accretion luminosity 
$L_{\rm acc} \equiv G M_* \dot{M}/R_*$ \cite{com1}.

\begin{figure}[h]
\begin{center}
\psfig{file=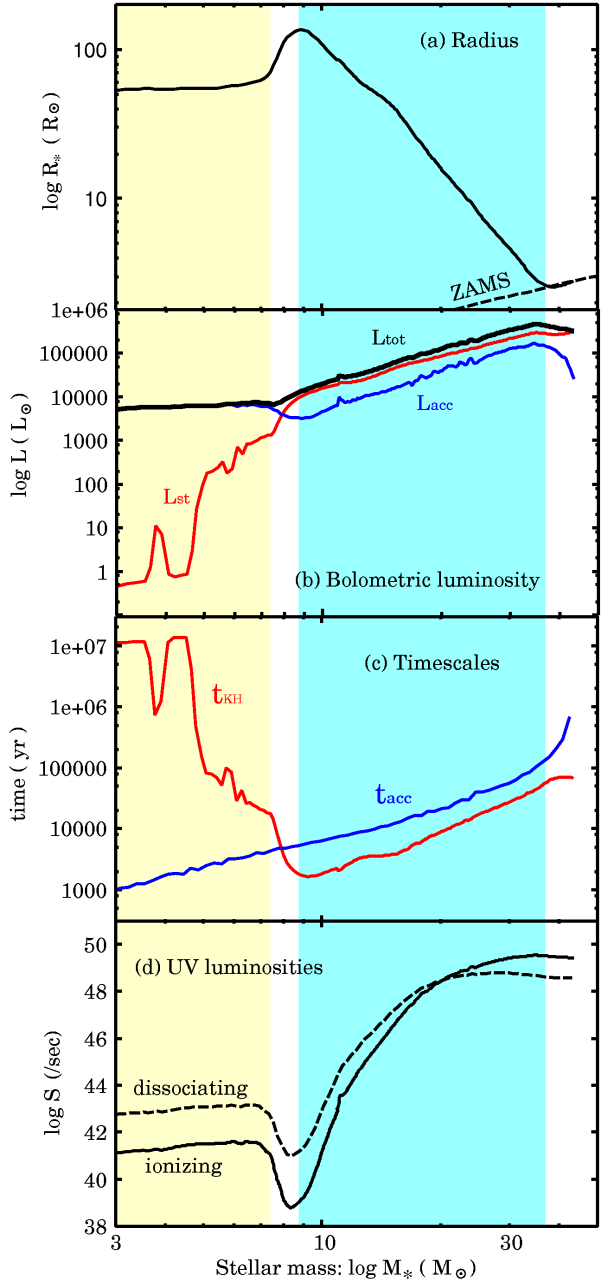,width=0.5\textwidth}
\caption{ Evolution of the stellar radius,
bolometric luminosity, evolutionary timescales, and 
ionizing ($h \nu \geq 13.6~{\rm eV}$) and 
dissociating ($11~{\rm eV} \leq h \nu \leq 13.6~{\rm eV}$) 
photon number luminosity.
In panel (b), the total luminosity $L_{\rm tot}$ (black)
is the sum of the stellar luminosity $L_*$ (red) 
and accretion luminosity $L_{\rm acc}$ (blue).
The KH timescale $t_{\rm KH}$ (red) and accretion 
timescale $t_{\rm acc}$ (blue) are depicted in panel (c).
The yellow and blue backgrounds denote the adiabatic accretion
phase and KH contraction phase in the protostellar evolution.}
\label{fig:pevol}
\end{center}
\end{figure}

In the early accretion phase, the stellar radius remained almost
constant at $\simeq 50$ solar radius ($R_\odot$) (Fig. \ref{fig:pevol}-A).
The stellar luminosity was substantially 
lower than the accretion luminosity (Fig. \ref{fig:pevol}-B) and 
the KH timescale was much longer than the accretion time 
(Fig. \ref{fig:pevol}-C). 
Consequently, entropy carried by the accreted gas
accumulated at the stellar surface nearly without loss.
During this quasi-adiabatic stage ($M_* < 7~M_\odot$),
the luminosity $L_{\ast}$ increased with stellar mass. 
When the star grew to $8~M_\odot$, 
the KH timescale finally fell below the accretion 
timescale (Fig. \ref{fig:pevol}-C). 
After this, the protostar began its so-called KH contraction, 
in which it gradually contracted as it radiated its energy away
(Fig. \ref{fig:pevol}-A).
The stellar luminosity was the main component of the total
luminosity after this evolutionary stage (Fig. \ref{fig:pevol}-B).
The stellar luminosity $L_*$ increased, and stellar radius $R_*$
decreased, as the stellar mass increased.
As a result, the stellar effective temperature 
$T_{\rm eff} \propto (L_*/R_*^2)^{1/4}$ and the ultraviolet (UV)
flux rapidly rose (Fig. \ref{fig:pevol}-D). 
Thus, ionization and heating by UV photons became 
important already in the KH contraction stage.

\begin{figure}
\begin{center}
\psfig{file=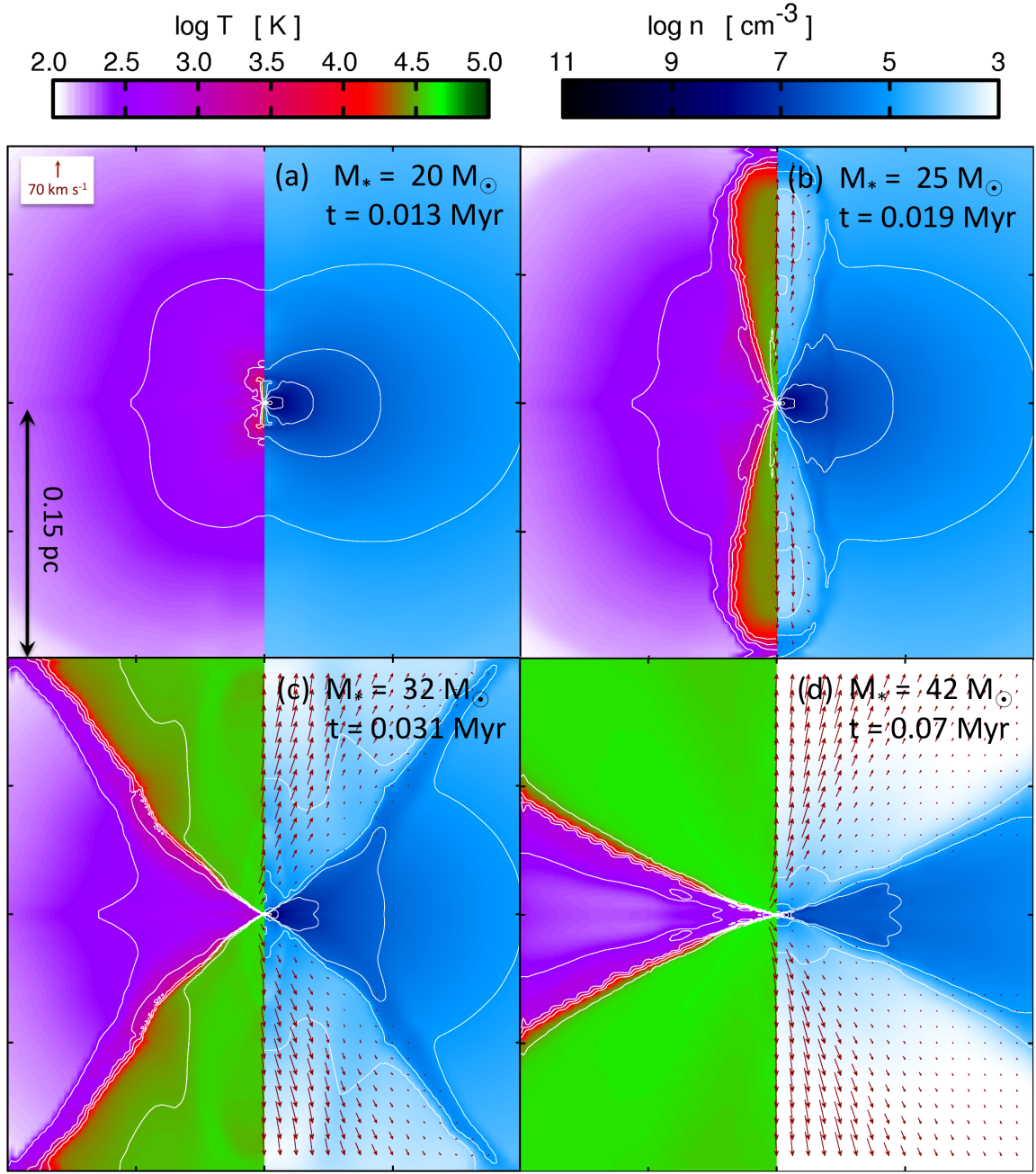,width=1\textwidth}
\caption{UV radiative feedback from the primordial protostar.
The spatial distributions of gas temperature 
({\it left}), number density ({\it right}), and velocity 
({\it right, arrows}) are presented in each panel for the central 
regions of the computational domain.
The four panels show snapshots at times, when the stellar mass is
$M_* = 20~M_\odot$ ({\it a}), $25~M_\odot$ ({\it b}), 
$35~M_\odot$ ({\it c}), and $42~M_\odot$ ({\it d}). 
The elapsed time since the birth of the primordial protostar 
is labeled in each panel. }
\label{fig:shots}
\end{center}
\end{figure}
\begin{figure}
\begin{center}
\psfig{file=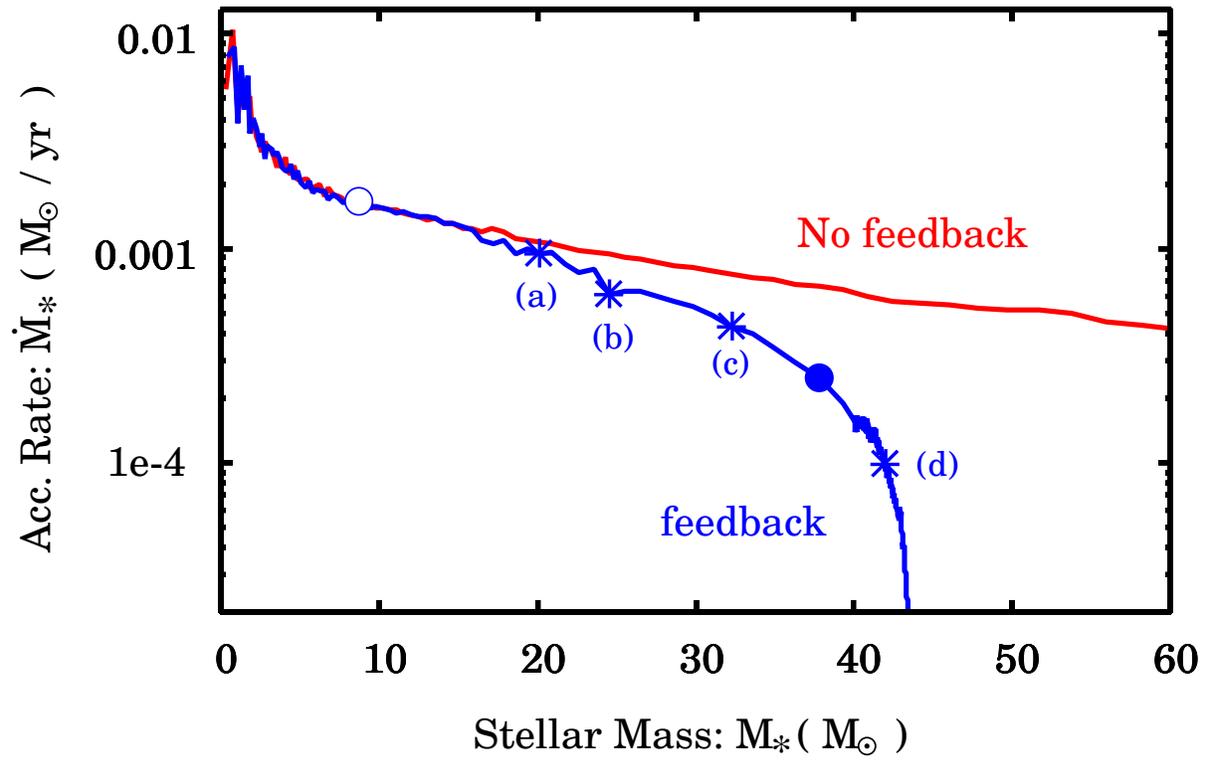,width=1\textwidth}
\caption{Evolution of the accretion rate onto the primordial
protostar. The blue line indicates the evolution, which includes 
the effect of UV radiative feedback from the protostar.
The red line indicates a numerical experiment with no UV feedback.
The open and solid circles denote the characteristic epochs
of the protostellar evolution, beginning of the KH contraction
and the protostar's arrival to the ZAMS.
Figure \ref{fig:shots}, A to D, shows the snapshots 
at the moments marked here with asterisks.}
\label{fig:xmdot}
\end{center}
\end{figure}

When the stellar mass was $20~M_\odot$, 
an ionized region rapidly expanded in a bipolar shape 
perpendicular to the disk, where gas was  
cleared away (Fig. \ref{fig:shots}-A).
At this moment, the disk extended out to $\simeq 600$~AU. 
The disk was self-shielded against the stellar H$_2$-dissociating 
($11.2~{\rm eV} \leq h \nu \leq 13.6~{\rm eV}$) as well as the
ionizing radiation. 
The ionized atomic hydrogen (HII) region continued to grow and 
finally broke out of the accreting envelope. 
At $M_* \simeq 25~M_\odot$, the size of the bipolar HII region 
exceeded 0.1~pc (Fig. \ref{fig:shots}-B). 
Because of the high pressure of the heated ionized gas, the opening angle 
of the ionized region also increased as the star grew 
(Fig. \ref{fig:shots}-C).
Shocks propagated into the envelope preceding the expansion of the 
ionized region. The shocked gas was accelerated outward 
at a velocity of several kilometers per second.
The shock even reached regions shielded against direct
stellar UV irradiation.
The outflowing gas stopped the infall of material from the envelope 
onto the disk (Fig.~S6). 
Without the replenishment of disk material from the envelope the 
accretion rate onto the protostar decreased (Fig.~\ref{fig:xmdot}).
In addition, the absence of accreting material onto the 
circumstellar disk means that the disk
was exposed to the intense ionizing radiation from the star.
The resulting photoevaporation of disk gas
also reduced the accretion rate onto the protostar. 
The photoevaporated gas escaped toward the polar direction 
within the ionized region. The typical velocity of the
flow was several tens of kilometers per second, 
comparable with the sound speed of the ionized gas, which was high enough
for the evaporating flow to escape from the gravitational potential
well of the dark matter halo.


When central nuclear hydrogen burning first commenced at a stellar 
mass of $35~M_\odot$, it was via the proton-proton (pp) -chain 
normally associated with low-mass stars.  
The primordial material does not have the nuclear
catalysts necessary for carbon-nitrogen-oxygen (CNO)
-cycle hydrogen burning.  Because
the pp-chain alone cannot produce nuclear energy at the rate
necessary to cover the radiative energy loss from the stellar surface,
the star continues to contract until central temperatures and
densities attain values that enable the 3-$\alpha$ process 
of helium burning \cite{OP03}. 
The product of helium burning is carbon, and once the relative mass
abundance of carbon reaches $\sim 10^{-12}$, CNO-cycle hydrogen burning
takes over as the principal source of nuclear energy production, 
albeit at much higher central densities and temperatures than those of 
stars with solar abundances. These first-generation ZAMS (zero-age main
sequence) stars are thus more compact and hotter than their 
present-day counterparts of equal mass \cite{EC71}.
The subsequent evolution of the accreting star followed along the
ZAMS mass-radius relationship (Figure \ref{fig:pevol}A).      
By the time the star attained $40~M_\odot$,
the entire region above and below the disk (Fig. \ref{fig:shots}D) 
was ionized. Mass accretion  was terminated when the stellar mass was  
$43~M_\odot$ (Fig. \ref{fig:xmdot}).


The entire evolution described above
took about $0.1$ million years from the birth
of the embryo protostar to the termination of the accretion.
The star is expected to live another few million 
years before exhausting all available nuclear fuel and
exploding as a core-collapse supernova\cite{Schaerer02}.


Our calculations show that the first stars regulated
their growth by their own radiation.
They were not extremely massive, 
but rather similar in mass to the O-type stars in our Galaxy.
This resolves a long-standing enigma regarding
the elemental abundance patterns of the Galactic oldest 
metal-poor stars, which contain nucleosynthetic signatures 
from the earliest generation of stars.
If a substantial number of first stars had masses in excess 
of $100~M_{\odot}$, they would end their lives through
pair-instability supernovae\cite{UN02,HW02}, expelling
heavy elements that would imprint a characteristic nucleosynthetic
signature to the elemental abundances in metal-poor stars. 
However, no such signatures have been detected in 
the metal-poor stars in the Galactic halo\cite{TVS04,FJB09}.
 For example, the abundances of elements with odd atomic numbers are
generally reduced in remnants of primordial supernovae.
The odd-even contrast pattern expected
in pair-instability supernovae is
much stronger than the observed patterns in Galactic 
metal-poor stars\cite{HW02}.
Moreover, pair-instability supernovae predict a small
abundance ratio [Zn/Fe], but observed values
are much larger\cite{UN02}.
Detailed spectroscopic studies of extremely metal-deficient
stars indicate that the metal-poor stars were born in 
an interstellar medium that had been metal-enriched
by supernovae of ordinary massive stars\cite{Iwamoto05}.


Second-generation stars, which formed from
the primordial gas affected by radiative or mechanical
feedback from the first stars, could have dominated the metallicity of
the young interstellar medium, which then spawned the observed Galactic
halo stars. 
These second-generation stars could have 
been more numerous but less massive than the first stars
because of a different gas thermal evolution, with additional 
radiative cooling via H$_2$ and HD molecules \cite{Osh05,Y07}.
However, this mode of star formation is suppressed 
even with weak H$_2$ photodissociating background radiation 
\cite{Y07}.
If so suppressed, the formation process of the later-generation 
primordial stars would be similar to that of the very first stars,
and most primordial stars could have
experienced the evolution presented in this article regardless
of their generation.
One might argue that today's observed metal-deficient stars formed 
after an episode of star formation with non-zero metallicity.  For the
star formation process to differ substantially from that of the very first
stars, one would require metallicities in excess of 
[Fe/H] $ > -5$\cite{com2,HO09b}. 
Caffau and co-workers \cite{Caffau11} report on observations of a
metal-deficient star with [Fe/H] $\simeq -5$ but without the corresponding
enhancement of carbon, nitrogen, and oxygen found in metal-deficient
stars. The abundance pattern of this star agrees with expectations
from core-collapse supernovae, implying that
it formed from gas enhanced by material ejected from
primordial stars with masses less than $100~M_\odot$.

We have performed radiation-hydrodynamic simulations
only for a single star-forming region embedded in a cosmological
simulation.  
Our selected dark halo was typical in mass, spin, and 
formation epoch, when compared with those in other studies
\cite{ON07}.
The evolution presented here is somewhat similar to that predicted
by the semi-analytic model, in which the expansion of the ionized 
region begins when the stellar mass is $\simeq 25~M_\odot$ 
and the final mass is $\simeq 57~M_\odot$ \cite{MT08},
the lowest final stellar mass predicted by
the semi-analytic treatment.
If input parameters of the semi-analytic models
are chosen to fit our initial gas cloud, however, the final mass should be 
higher, $\simeq 90~M_\odot$ (SOM text). 
Our calculations follow the dynamical response of the infalling gas
onto the circumstellar disk.
The expansion of the ionized region around the protostar generates a 
powerful outflow even behind the surrounding disk. This effect 
reduces the accretion rate significantly.
Thus, our radiation-hydrodynamic calculations predict
systematically lower final masses than those of the semi-analytic models.

Although the results described above provide
a complete picture of how a primordial protostar regulates and
terminates its growth, there are a few key quantities that
determine the strength of the feedback effect,
as suggested by the semi-analytic model\cite{MT08}.
With smaller initial rotation of the natal dense core, the
density in the envelope near the polar directions 
would be higher. Hydrogen recombination occurs rapidly
in the dense gas, which prevents the breakout of the ionized region. 
Gas accretion can last for a longer time in this case, 
and would form more massive stars (SOM text).
Nevertheless, even considering variations among dark halos bearing the first
stars, a substantial fraction of the first stars should
be less massive than $100~M_\odot$ and end their lives as
ordinary core-collapse supernovae.
Gas accretion might not be completely halted in a few exceptional
cases, thus leading to the formation of a small number of extremely massive 
stars that are $> 100~M_\odot$ in the early universe\cite{Ohkubo09}.
Black holes left after such very massive stars' deaths might have
grown up to be the supermassive black holes lying in galaxies.

Recent 3D cosmological simulations 
showed that a primordial gas cloud breaks up into several 
embryo protostars in an early phase\cite{TAO09,Cl11}. 
Each of these protostars would continue to grow through mass accretion,
but it remained uncertain how and when the growth is
halted. Our radiation-hydrodynamics calculations 
explicitly show that the parent gas cloud is evaporated 
by intense radiation from the central star when 
its mass is several tens of solar-masses.
The circumstellar disks in our simulations
are marginally stable against gravitational fragmentation [Fig.~S5].
We expect that these disks -- in a 3D simulation -- would evolve
analogously to our numerical simulations with assumed axial symmetry and
have a similar time-averaged structure.  
It is conceivable that a few protostars are ejected 
dynamically from the parent cloud, to remain as low-mass stars.
Observationally, however, there have been no low-mass 
zero-metallicity stars discovered in the Galaxy. 
This fact suggests the limited formation efficiency of such
low-mass primordial stars\cite{FJB09}.
Low-mass stars ($< 1~M_\odot$) and extremely high mass stars 
($>100~M_\odot$), if any, are thus a minor population 
among the first stars. 

Our self-consistent calculations show that 
the characteristic mass of the first stars is
several tens of solar masses.
Although this is less than that of the conventionally
proclaimed extremely massive stars ($> 100~M_{\odot}$), 
it is still much larger than the characteristic mass of stars 
in our galaxy ($< 1~M_{\odot}$)\cite{C03}.

\bibliography{biblio}

\begin{thebibliography}{10}
\bibitem{com1}
{This definition of accretion luminosity includes the mechanical luminosity of
  an accretion-driven wind and the ratio $t_{\rm KH}/t_{\rm acc}$ is equal to
  the ratio $L_{\rm acc}/L_*$} .
\bibitem{com2}
{$[{\rm Fe/H}] \equiv \log ({\rm Fe/H})_{\rm star} - \log ({\rm Fe/H})_{\rm
  sun}$, where $({\rm Fe/H})$ is mass ratio of Fe to H} .
\end{thebibliography}
\bibliographystyle{Science} 

{\bf Acknowledgments:}
We thank T. Nakamura, K. Nomoto, S. Inutsuka, and N. Turner for 
stimulating discussions on this topic. Comments by an anonymous
referee helped improve the manuscript.
T.H. appreciates the support by Fellowship of the Japan
Society for the Promotion of Science for Research Abroad.
The present work is supported in part by the grants-in-aid
by the Ministry of Education, Science and Culture of Japan
(19047004, 2168407, 21244021:KO, 20674003:NY).
Portions of this research were conducted at the Jet 
Propulsion Laboratory, California Institute of Technology, which is 
supported by NASA.
Data analysis was (in part) carried out on the general-purpose PC 
farm at Center for Computational Astrophysics (CfCA) of National 
Astronomical Observatory of Japan.

\vskip 1cm
\noindent
{\bf Supporting Online Material}\\
www.sciencemag.org\\
SOM text\\
Figures S1-S7\\
Tables S1 and S2\\
References (30-64) \\
\\
25 April 2011; accepted 28 October 2011 \\
Published online 10 November 2011
                   


\clearpage

\begin{center}
{\LARGE {\bf Supporting Online Material}} 
\end{center}

\section{Numerical Methods}

We have developed a hybrid code to simultaneously
calculate the evolution of an accreting first generation
(proto-)star and the flow of the surrounding
primordial material.
We solve the accretion flow with a radiation hydrodynamic
code, whereby the central (proto-)star is represented as a 
sink cell. This sink cell grows in mass by inflow through its 
boundary and it influences the surrounding material flow via 
gravity and by emitting radiation. 
We follow the evolution of the central star from the earliest 
protostellar stage up to the early phases of nuclear burning on 
the zero age main sequence by solving the conventional stellar 
structure equations, taking into account mass accretion.  The material
functions for the stellar evolution calculation -- the opacity, 
the equation of state and other thermodynamic relationships, 
as well as the nuclear reaction network -- assume zero metallicity.
The calculated evolution provides the emitted radiation 
from the sink cell as a function of time. 
We describe below our methods of calculation for the accretion flow 
and the protostar, respectively.  

\subsection{Radiation Hydrodynamics of the Accretion Flow}

To study the evolution of the flow of primordial gas we employ
a grid-based axisymmetric radiation hydrodynamic
code with self-gravity, previously used for studying
present-day star formation {\it (7,10)} and the evolution of
photoionized gas flow from protostellar disks 
\cite{YW96,RY00}.
This code makes use of a nested-grid technique to cover a wide 
spatial range \cite{YK95}.  We have added a chemical network 
besides changes based on the nature of the primordial material.

The governing equations for gas dynamics in cylindrical 
coordinate $(R,Z)$ are,
\begin{equation}
\frac{\partial \rho}{\partial t} + \nabla \cdot (\rho \vvec) = 0 ,
\end{equation}
\begin{equation}
\frac{\partial (\rho \vvec)}{\partial t} 
+ \nabla \cdot ( \rho \vvec \otimes \vvec ) =
- \rho \nabla \Phi - \nabla p + \frac{A^2}{\rho R^3} \nvec_R 
+ \Kvec
\end{equation}
\begin{equation}
\frac{\partial e}{\partial t} + \nabla \cdot (e \vvec)
= - p \nabla \cdot \vvec + \Gamma - \Lambda ,
\end{equation}
\begin{equation}
p = (\gamma - 1) e ,
\end{equation}
where $\rho$ and $p$ are the gas density and pressure, $\Phi$
the gravitational potential, 
$\vvec$ the 2-dimensional (2D) velocity vector $(v_R,v_Z)$, 
$A \equiv \rho R v_\phi$ the angular momentum per unit volume,
$\nvec_R$ the radial unit vector,
$\Kvec$ the radiation force,
$e$ the gas internal 
energy density, $\Gamma$ and $\Lambda$ the heating and cooling 
rates per unit volume, and $\gamma$ the adiabatic exponent.
We calculate $\gamma$ from the chemical composition
at each grid cell as in {\it (2)}.
The radiative and chemical processes included in 
the energy source term $\Gamma - \Lambda$ are summarized in Table S1.

\begin{table}
\begin{center}
Table S1. Included thermal processes \\[3mm]
\begin{tabular}{llc}
\hline
\hline  & Processes & References \\
\hline \\
        & Photoionization   & \cite{OF06} \\
Heating & Photodissociation & \cite{HM79} \\
        & H$_2$ formation   & \cite{HM79}, \cite{SK87}  \\
\\
\hline
\\
        & H$_2$ collisional excitation & \S~\ref{sssec:rad} \\
        & H$^-$ free-bound emission & \S~\ref{sssec:rad} \\
        & H$_2$ collisional dissociation & \cite{HM79}, \cite{SK87} \\
Cooling & H collisional ionization & \cite{OF06} \\
        & H collisional excitation &  \cite{An97} \\   
        & Compton scattering &  \cite{An97} \\
        & HeII collisional excitation & \cite{An97} \\
\\
\hline
\end{tabular}
\end{center}
\label{tab:thprocess}
\end{table}

\subsubsection{Radiation Transfer}
\label{sssec:rad}

For the problem at hand photons in different wavelength regimes 
play vastly different roles in the thermal and chemical processes.
For example, whereas gas cools via molecular hydrogen line emission
and infrared/optical continuum radiation, photons with energies
$h \nu \geq 13.6$~eV (EUV) photoionize atomic hydrogen and 
photons with energies $11.2~{\rm eV} \leq h \nu \leq 13.6~{\rm eV}$ (FUV) 
photodissociate molecular hydrogen.
To treat their transfer properly
without time-consuming calculations, we adopt different 
methods for each of the radiation components.

\paragraph{Molecular Hydrogen Line Cooling}

The primary cooling process in low-temperature 
($< \sim 8000$K) primordial gas 
is line emission via rotational and vibrational transitions 
of hydrogen molecules.
We adopt the fitting formula by \cite{HM79} for the optically-thin
limit for $n < \sim 10^9~{\rm cm}^{-3}$.
Dense gas with $n > \sim 10^9~{\rm cm}^{-3}$ is opaque 
to H$_2$ line emission, and the cooling rate is reduced 
by photon trapping.
In this case the cooling rate is calculated by adapting 
the method of \cite{Y06} 
to the case of axial symmetry.
We sum up cooling rates for all possible transitions among the 
rotational levels from $J=0$ to 20 
and vibrational levels $v=0,1,2$.
The cooling rate by a transition for the optically thick case 
is obtained by multiplying the value for optically thin 
cooling by the escape probability.  
The escape probability $\beta_{\rm esc}$ is evaluated as
\begin{equation}
\beta_{\rm esc} = \frac{\beta(\tau_R) + \beta (\tau_Z)}{2} ,
\end{equation}
where $\beta (\tau)$ is
\begin{equation}
\beta (\tau) = \frac{1 - \exp (\tau)}{\tau}, 
\end{equation}
and $\tau_R$ and $\tau_Z$ are the optical depths 
of this transition along $R$ and $Z$ directions, respectively. 
We calculate the optical depths using the local velocity gradients
(so-called Sobolev approximation)
\begin{equation}
\tau_{q} = \alpha \frac{c_s}{| \partial v_q / \partial q |} ,
\end{equation}
where $q=R$ or $Z$, $\alpha$ is the absorption coefficient 
and $c_s$ is the sound speed.

For calculating the level-population of H$_2$ molecules, we omit the
excitation by absorbing stellar photons, which could potentially 
reduce the line cooling rate. 
In general, H$_2$ molecules could be excited to (i) 
the Lyman-Werner bands by absorbing FUV photons 
or (ii) the higher rotational and vibrational
levels by absorbing infrared photons. 
In our case, the process (i) would be minor because the disk
is shielded against the stellar FUV radiation (also see Sec.~\ref{sec:disk}). 
The process (ii) is negligible in most parts of the disk
even with the radiation from the protostar.

\paragraph{Continuum Cooling}
During the collapse phase before the formation of a protostar, 
continuum cooling via H$_2$ collision-induced emission 
(CIE) is important in the dense ($n > 10^{13}~{\rm cm}^{-3}$) 
molecular gas {\it (2)}. 
However, we only find such dense gas inside the sink cell;
it does not otherwise appear in our hydrodynamical calculation.
Instead, cooling via H$^-$ free-bound emission becomes 
important in the nearly vertical flows onto the circumstellar 
disk (also see Sec.~\ref{sec:disk} below). 

The continuum cooling rate $\Lambda_c$ in the gray approximation
is written as
\begin{equation}
\Lambda_c =  c \rho \kappa_P ( 4 \pi B(T_g) - E_c) ,
\end{equation}
where $c$ is the speed of light, and
$\kappa_P$ is the Planck mean opacity per unit mass.
The continuum radiation energy density $E_c$ 
satisfies the equation
\begin{equation}
\frac{\partial E_c}{\partial t} 
= - \nabla \cdot \Fvec_c + \Lambda_c + j_* ,
\label{eq:ecrad}
\end{equation}                                               
where $\Fvec_c$ is the energy flux, and $j_*$ is the stellar source term.
Using the total luminosity of the protostar $L_{\rm tot}$, 
the stellar source term is given by 
\begin{equation}
j_* = \frac{L_{\rm tot}}{2 \pi \Delta R^2 \Delta Z},
\end{equation}
where $\Delta R$ and $\Delta Z$ are the size of the sink cell,
which is equal to the cell size at the finest grid-level, 12~AU.
The source term is zero except for the sink cell.
We solve equations (\ref{eq:ecrad}) with the flux-limited 
diffusion (FLD) approximation \cite{LP81} using 
the operator splitting technique with $de/dt = - \Lambda_c$
\cite{TS01}.
The FLD method adopts a closure relation between $\Fvec_c$ 
and $E_c$ 
\begin{equation}
\Fvec_c = - \frac{c \lambda}{\rho \kappa_R} \nabla E_c ,
\label{eq:fldc}
\end{equation}
where $\kappa_R$ is the Rosseland mean opacity,
and $\lambda$ is the flux-limiter defined as
\begin{equation}
\lambda = \frac{2 + s}{6 + 3s + s^2} ,
\end{equation}                         
\begin{equation}
s = \frac{ | \nabla E_c |}{E_c \rho \kappa_R} .
\end{equation}
We use the opacities $\kappa_P$ and $\kappa_R$
for primordial gas calculated by \cite{MD05} in tabulated forms.
As mentioned above, the most important continuum cooling is 
that via H$^-$ free-bound emission. 
Neutral gas falling onto the disk is 
heated up to $T_g \simeq 5 \times 10^3$~K by compressional heating,
until thermal balance is achieved with this cooling process. 
In our calculation the accreting envelope is always 
optically thin to the non-UV continuum radiation and non-UV stellar radiation 
escapes without significant heating in the envelope. 

\paragraph{Stellar EUV and FUV Radiation}

EUV radiation from the star ionizes the material in its immediate vicinity.
Within the HII region, there are two components of the EUV field:
the direct stellar component and the
diffuse component emitted via recombinations directly into the 
ground state (i.e. no ``on-the-spot'' approximation).
We solve the transfer of these two EUV components separately
following \cite{RY00}.
We adopt a frequency-averaged approximation for each component, 
taking into account the difference in their mean energies,
$h \bar{\nu}_*$ (stellar) and $h \bar{\nu}_d$ (diffuse).

To calculate the direct EUV field we 
cast a number of radial rays from the central star to the outer edge of the 
simulation box, along which the transfer equation for 
the direct EUV photon number flux $\FFvec_*$ is solved:
\begin{equation}
\nabla_r \cdot \FFvec_* = - n (1 - x) \sigma_* \FFvec_* ,
\label{eq:euvst}
\end{equation}
where $x$ is the degree of ionization and $\sigma_*$ is the absorption 
cross section per particle.
The cross section $\sigma_*$ is a function of mean energy of
the direct EUV photons, i.e., the effective temperature of the star
$T_{\rm eff}$ \cite{OF06}.

We employ the FLD approximation for the diffuse component, whose 
transfer equation is
\begin{equation}
\frac{\partial {\cal N}_d}{\partial t} = 
- \nabla \cdot \FFvec_d + \alpha_1 (T_g) n^2 x^2 
- n (1-x) \sigma_d c {\cal N}_d ,
\label{eq:euvd}
\end{equation}
where ${\cal N}_d$ is the diffuse EUV photon number density, 
$\alpha_1$ is the recombination coefficient to the ground state,
and $\sigma_d$ is the absorption cross section.
Under the approximation that the mean energy of the diffuse component
depends only on the local gas temperature $T_g$, the cross 
section $\sigma_d$ becomes a function of $T_g$ \cite{OF06}.
In the FLD approximation,  
\begin{equation}
\FFvec_d = - \frac{c \lambda}{n (1-x) \sigma_d} \nabla {\cal N}_d , 
\end{equation}                                                       
in analogy to equation (\ref{eq:fldc}).
The photoionization heating and recombination cooling rates are
calculated for each component of EUV radiation separately.

As for the photodissociating FUV radiation, 
we only consider the stellar direct component.
We calculate the photodissociation rate by multiplying 
the value for the optically-thin case by the self-shielding factor 
\cite{DB96}
\begin{equation}
f_{\rm sh} (N_{\rm H_2})= \left\{
\begin{array}{ll}
1 & (N_{\rm H_2} < N_1) \\
(N_{\rm H_2}/N_1)^{-3/4} &
(N_1 < N_{\rm H_2} < N_2) \\
0 & (N_{\rm H_2} > N_2) ,
\end{array}
\right.
\label{eq:fsld}
\end{equation}
where $N_{\rm H_2}$ is the H$_2$ column density 
from the protostar to the point under consideration,   
$N_1 = 10^{14}~{\rm cm}^{-2}$, and $N_2 = 10^{22}~{\rm cm}^{-2}$.
Although the original functional form of 
$f_{\rm sh}(N_{\rm H_2})$ by \cite{DB96} is for the range 
$N_1 < N_{\rm H_2} < N_2$, 
we adopt the cut-off at $N_{\rm H_2} > N_2$, 
where the photodissociation rate falls sharply 
according to \cite{DB96}. 
In our calculation, the column density largely exceeds 
$N_2$ along the radial rays incident on the circumstellar disk.

As shown in Sec.~\ref{sec:disk} below, the innermost part of the disk
at $R <$ several 10~AU is not spatially resolved in our 
calculations.
We expect the innermost part of the disk to be optically thick
and shield the material near the disk mid-plane behind it 
from the stellar radiation \cite{TM04}. 
We adopt the following treatment for modeling this effect. 
During the calculation, we evaluate the disk scale height 
$H (R) \equiv c_s/\Omega$ at each radius $R$ on the equator.
We compute $i_{R,d}$, the $R$-index of the cell at the outer edge
of the unresolved part, for which $H (R) < \Delta Z$.
The unresolved part is assumed to be opaque to radial rays incident
on the $(i_{R,d}+1)$-th cell on the equator up to a height $\Delta Z$,
but transparent for the other rays.

\paragraph{Radiation Force}

The total radiation force $\Kvec$ is calculated as the sum 
of contributions by the continuum radiation, 
direct and diffuse EUV radiation:
\begin{equation}
\Kvec =   \frac{\rho \kappa_R}{c} \Fvec_c 
        + \frac{n (1-x) \sigma_*}{c} h \bar{\nu}_* \FFvec_*
        + \frac{n (1-x) \sigma_d}{c} h \bar{\nu}_d \FFvec_d  .
\end{equation}
Here, we omit contributions from FUV photons. 
In reality, the gas accretion envelope is opaque against
Lyman-$\alpha$ photons emitted from the stellar atmosphere and
HII region. Radiation pressure via the Lyman-$\alpha$ scattering 
is exerted on the accretion envelope.
With mass accretion through the circumstellar disk, in particular, photons
are preferentially transferred toward the polar direction, where the gas
density rapidly decreases as the mass accretion proceeds
(``flashlight effect'': c.f. {\it (7)}). The semi-analytic modeling 
by {\it (8)} shows that the Lyman-$\alpha$ pressure could influence the 
dynamics of infalling material near the rotation axis.
As also discussed in {\it (8)}, however, the radiation pressure 
via Lyman-$\alpha$ scattering would be significantly reduced once
gas in polar directions is blown away and photons 
escape from the cavity. In this paper, we focus on the UV radiative 
feedback effects to derive upper limits of the stellar final masses.

\subsubsection{Chemical Reactions in the Primordial Gas}  

We consider the five species of primordial gas
H, H$^+$, e, H$_2$, and H$^-$,  
based on the minimal model of \cite{Abel97} with some
additional reactions.
The included reactions are summarized in Table S2.
Except H$^-$, we solve kinetic equations  
with an implicit difference scheme.
H$^-$ is assumed to be in chemical equilibrium.
Photodissociation of H$^-$ is omitted in the adopted 
chemical network.
In our calculations the gas density is high enough that the 
collisional destruction processes (R4, 14, and 15) control 
its abundance. We curtail helium chemistry by  
assuming that helium is singly ionized in an HII region 
and atomic elsewhere.
We do not include deuterium reactions because HD
cooling is relevant only in low-temperature ($T_g < 200$~K)
and low-density ($n < 10^8{\rm cm^{-3}}$) gas, 
which does not appear in our simulations.

\begin{table}[t]
\begin{center}
Table S2. Included chemical reactions \\[3mm]
\begin{tabular}{llc}
\hline
\hline No. & Reactions & References \\
\hline \\
R1 &  H      +  e  $\rightarrow$ H$^+$ + 2 e       &  \cite{Abel97} \\
R2 &  H$^+$  +  e  $\rightarrow$ H     + $\gamma$  &  \cite{OF06}   \\
R3 &  H$^-$  +  H  $\rightarrow$ H$_2$ + e         &  \cite{GP98}  \\
R4 &  H$_2$  +  H$^+$ $\rightarrow$ H$_2^+$ + H    &  \cite{GP98}  \\
R5 &  H$_2$  +  e     $\rightarrow$ 2 H + e        &  \cite{GP98}  \\
R6 &  H$_2$  +  H  $\rightarrow$  3 H              &  \cite{SK87}  \\
R7 &  3 H          $\rightarrow$  H$_2$ + H        &  \cite{PSS83} \\
R8 &  2 H  + H$_2$ $\rightarrow$  2 H$_2$          &  \cite{PSS83} \\
R9 &  2 H$_2$      $\rightarrow$  2 H   + H$_2$    &  \cite{PSS83} \\
R10 & H    + e     $\rightarrow$  H$^-$ + $\gamma$ &  \cite{GP98}  \\  
R11 & 2 H          $\rightarrow$  H$^+$ + e + H    &  \cite{PSS83} \\
R12 & H + $\gamma$ $\rightarrow$  H$^+$ + e        &  \cite{OF06}   \\
R13 & H$_2$ + $\gamma$ $\rightarrow$ 2 H           
    &  \cite{TH85} and (\ref{eq:fsld})   \\  
R14 & H$^-$ + e    $\rightarrow$  H + 2 e          &  \cite{Abel97} \\
R15 & H$^-$ + H$^+$ $\rightarrow$ 2 H      &  \cite{GP98}   \\
\\
\hline
\end{tabular}
\end{center}
\label{tab:reactions}
\end{table}

\subsubsection{Angular Momentum Transport in the Accretion Disk}
\label{sssec:ang}

In an exactly axisymmetric system, angular momentum must be conserved.
In reality, however, a massive circumstellar disk 
can develop a non-axisymmetric spiral pattern, 
which exerts a torque on the matter in the disk 
and transfers angular momentum outward. 
Recent 3-dimensional (3D) numerical simulations demonstrated that 
this mechanism operates in fact in the circumstellar disks of the 
first stars \cite{SGB10,Cl11}.
We mimic this effect by adopting the angular 
momentum transport via the so-called $\alpha$-viscosity \cite{SS73}. 
The equation of angular momentum transport is thus given by
\begin{equation}
\frac{\partial A}{\partial t} + \nabla \cdot (A \vvec) 
= -  \frac{1}{R} \frac{\partial}{\partial R}
                 \left( R^3 \eta \frac{\partial \Omega}{\partial R} \right),
\end{equation}                                                    
where $\Omega$ is angular velocity, 
$\eta = 2 \alpha \rho c_s^2 / ( 3 \Omega )$, 
and $\alpha$ is a dimensionless free parameter. 

We assume that the $\alpha$-parameter depends on the height from the equator, 
\begin{equation}
\alpha(R,Z) = \alpha_0  \exp \left( - \frac{Z}{H (R)} \right) ,
\label{eq:alpha}
\end{equation}
where $\alpha_0$ is a constant. 
The effective values of the $\alpha$-parameter in rapidly accreting 
circumstellar disks are $\alpha \simeq 0.1 - 1$ as estimated from 
3D simulations of present-day massive star formation 
\cite{Krum09} as well as the formation of the first stars {\it (28)}.
In the fiducial case explained in the main article, we adopted 
$\alpha_0 = 0.6$.

Figure \ref{fig:xmdot_a} shows the evolution of accretion rates
onto the protostar for different values of $\alpha_0$. 
Although the evolution is qualitatively similar in all cases,  
with higher $\alpha_0$ angular momentum is transferred more rapidly 
and therefore accretion rates onto the protostar become higher.
We see that the final stellar mass increases with $\alpha_0$. 
With the low value of $\alpha_0 = 0.3$, the final stellar mass is 
about $35~M_\odot$. This is close to the low-mass limit of $30~M_\odot$ 
predicted by {\it (3)}, who expected that this amount of gas would 
accrete onto the protostar in a few thousand years, which is much
shorter than the timescale of the protostellar evolution.
Even with $\alpha_0$ as large as unity, on the other hand, 
the final mass reaches at most $M_{\ast} \simeq 50~M_\odot$.
Note, however, that this does not exclude the possibility of 
formation of higher-mass stars. 
The semi-analytic models predict that more massive stars would 
form with weaker initial rotation of the natal core {\it (8)}
(also see Sec.~\ref{sec:mt08}).
Although rare, stars exceeding $100~M_\odot$ might still form 
in such circumstances.

Figure \ref{fig:xmdot_a} also shows the evolution in such 
a test case, whereby the initial angular momentum is artificially
reduced to 30\% of the original 
value.\footnote{The angular momentum is reduced at the beginning of
the 2D calculation, e.g., at a point when the central density is
$\simeq 10^6~{\rm cm}^{-3}$ in the run-away collapse 
stage (see Sec.~\ref{sec:setup}).}
The final stellar mass is $\simeq 85~M_\odot$ in this case.
The increase of the final mass is understood as follows. 
First, the gas density remains high in the accretion envelope in the 
polar directions for the same stellar mass.
The stellar EUV photons are consumed more effectively by photoionization
of neutral hydrogen generated via rapid recombination, which delays growth
of an HII region.
Second, the protostellar evolution differs from the fiducial
case, because of the higher accretion rates.
Figure \ref{fig:pevol_a} shows that, with the weaker rotation, 
the Kelvin-Helmholtz contraction stage is shifted to higher 
stellar masses. Correspondingly, the stellar EUV luminosity increases 
rapidly at a stage when the protostar is more massive than for the
fiducial case.
Because of these effects, the stellar UV feedback becomes effective
at higher stellar masses. 
We note that for all examined cases, the mass accretion ceases 
soon after the protostar's arrival to the zero age main sequence.

\subsection{Protostellar Evolution}

We follow the evolution of the central protostar by solving
the four stellar structure equations taking account of 
mass accretion \cite{OP03, HO09, HYO10}:
\begin{equation}
\left( \frac{\partial r}{\partial M} \right)_t = \frac{1}{4 \pi \rho r^2},
\label{eq:con} 
\end{equation}
\begin{equation}
\left( \frac{\partial P}{\partial M}  \right)_t = - \frac{GM}{4 \pi r^4}, 
\label{eq:mom}
\end{equation}
\begin{equation}
\left( \frac{\partial L}{\partial M} \right)_t 
= \epsilon - T \left( \frac{\partial s}{\partial t}  \right)_M ,
\label{eq:ene}
\end{equation}
\begin{equation}
\left( \frac{\partial s}{\partial M} \right)_t
= \frac{G M}{4 \pi r^4} \left( \frac{\partial s}{\partial p} \right)_T
  \left( \frac{L}{L_s} - 1  \right) C ,
\label{eq:heat}
\end{equation}
where $M$ is the Lagrangian mass coordinate, $\epsilon$ is the energy
production rate by nuclear fusion, $s$ is the specific entropy,
and $L_s$ is the radiative luminosity with adiabatic temperature
gradient.
The coefficient, $C$ in equation (\ref{eq:heat}) is unity if 
$L < L_s$ (i.e., in radiative layers), and given by the 
mixing-length theory if $L > L_s$ (i.e., in convective layers). 
We also solve the structure of the accretion flow 
inside the sink cell under the assumption of 
steady state and spherical symmetry.
The entire structure of both the protostar and accretion flow
is consistently determined to satisfy the jump 
conditions for the accretion shock at the stellar surface \cite{SST80}.
Mass accretion rates onto the protostar $\dot{M}_*$ are given
by mass inflow rates through the surface of the sink cell 
in the radiation-hydrodynamics calculation. 
With a high accretion rate of 
$\dot{M}_* > 10^{-4}~M_\odot~{\rm yr}^{-1}$, 
the accretion flow just above the stellar surface becomes opaque 
to the stellar radiation under the assumption of perfect spherical symmetry. 
The photosphere is located far from the stellar surface,
its temperature is reduced to $\simeq 6000$~K 
even for very massive stars of $M_* \simeq 100~M_\odot$ {\it (11)}.
With the realistic disk accretion, however,
the stellar surface at high latitude is not totally embedded in such 
an opaque flow and the UV radiation can freely radiate.
For this reason, we evaluate the protostellar EUV and FUV photon number 
luminosities using the fluxes at the stellar surface as
\begin{equation}
S_{\rm EUV} = 4 \pi R_*^2 \int_{\rm 13.6eV}^\infty
                   \frac{\pi B(T_{\rm eff})}{h \nu}~d \nu ,
\label{eq:sev}
\end{equation}
\begin{equation}
S_{\rm FUV} = 4 \pi R_*^2 \int_{\rm 11.2eV}^{\rm 13.6eV}
                   \frac{\pi B(T_{\rm eff})}{h \nu}~d \nu ,
\end{equation}                                                         
where $R_*$ is the stellar radius, $B(T_{\rm eff})$ is the Planck
function, and $T_{\rm eff}$ is defined as
\begin{equation}
T_{\rm eff} = \left( \frac{L_* + L_{\rm acc}}{4 \pi \sigma R_*^2} 
              \right)^{1/4} ,
\label{eq:teff}
\end{equation}
where $L_*$ is the stellar luminosity, 
$L_{\rm acc} \equiv G M_* \dot{M}_* / R_*$ is the accretion luminosity, 
and $\sigma$ is Stefan-Boltzmann constant.

The stellar UV luminosities given by equations (\ref{eq:sev}) 
- (\ref{eq:teff}) assume that all the gas reaching the stellar
surface releases its gravitational energy at the stellar surface.
In reality, some fraction of the gravitational energy 
would be radiated away from the circumstellar disk with a
lower effective radiation temperature than that of
the stellar surface.
Because the accretion luminosity $L_{\rm acc}$ is
only a few $10$~\% of the stellar luminosity $L_*$, when
the protostellar feedback begins to influence the mass accretion 
(see the main article), we do not expect a significant change in the overall
accretion process.

The stellar EUV photons are consumed mostly by photoionizing the recombined
hydrogen atoms within the HII region. As the recombination rate is
proportional to the square of density, this consumption is most
significant in the vicinity of the protostar. 
In our calculations, however, the flow structure very near the
protostar is masked by our assumption of a sink cell.
We evaluate the EUV consumption rate within the sink cell 
as follows.
First, we consider the spherical ``evacuation zone'', whose radius 
$R_{\rm ev}$ is smaller than the gravitational radius for the ionized gas,
\begin{equation}
R_{g,{\rm HII}} \equiv \frac{G M_*}{c_{s,{\rm HII}}^2}
\simeq 100~{\rm AU} \left( \frac{M_*}{10~M_\odot} \right)
                    \left( \frac{T_{\rm HII}}{10^4~{\rm K}} \right)^{-1} ,
\label{eq:rghii}
\end{equation}
and larger than the size of the sink cell, $\simeq 10$~AU.
In our calculations the temperature of ionized gas $T_{\rm HII}$ is
$\simeq 3 \times 10^4~K$ just after formation of the HII region and 
rises slightly as the stellar mass increases.
We adopt $R_{\rm ev} = 30$~AU as a fiducial value.
The density distribution within the evacuation zone should be well
approximated as the free-fall flow whose radial density structure
follows $\rho(r) \propto r^{-3/2}$.  The consumption rate of EUV
photons within this zone is analytically written as,
\begin{equation}
S_{\rm EUV, ev} = \frac{\alpha \dot{M}_{\rm ev}^2}{8 \pi \mu^2 G M_*}
                  \ln \left( \frac{R_{\rm ev}}{R_*}  \right) ,
\label{eq:seuv}
\end{equation}
where $\alpha$ is the total recombination rate, 
$\mu = (1 + 4 y_{\rm He}) m_p$ with the helium abundance $y_{\rm He}$
and proton mass $m_p$, and $\dot{M}_{\rm ev}$
is the mass inflow rate into this zone.
We evaluate $\dot{M}_{\rm ev}$ using the gas inflow rate 
along the $Z$-axis, 
\begin{equation}
\dot{M}_{\rm ev} = 4 \pi \rho_{z,{\rm ev}} R_{\rm ev}^2
                   \sqrt{ \frac{2 G M_*}{R_{\rm ev}} } ,
\end{equation}
where $\rho_{z,{\rm ev}}$ is the gas density at $(R,Z) = (0,R_{\rm ev})$.
 We suppose that the HII region is quenched inside the evacuation
zone when the stellar EUV luminosity $S_{\rm EUV}$ is less than the 
consumption rate $S_{\rm EUV,ev}$.
We do not solve the EUV radiative transfer for such cases. 
After $S_{\rm EUV}$ exceeds $S_{\rm EUV,ev}$, we remove the limit
of $S_{\rm EUV,ev}$ and assume that the EUV luminosity from the sink
is $S_{\rm EUV}$. 

We only marginally resolve the evacuation zone with the current
grid resolution.
The estimated value of $S_{\rm EUV,ev}$ depends on 
the grid size without resolving $R_{g,{\rm HII}}$,
because free-fall flow is valid only inside of $R_{g,{\rm HII}}$. 
For test calculations with a 2$\times$ coarser innermost grid around the 
protostar, formation of the HII region begins for a bit higher 
stellar mass (by a few $10\%$).
Moreover, the flow structure within the evacuation zone could be
complex. For instance, the protostellar outflow might be launched from 
the innermost part of the disk \cite{Mcd06} with the help of magnetic field
generated by dynamo amplification \cite{TB04, SL06}.
This outflow would help the breakout
of the HII region by clearing out materials close to the star
in the polar directions. However, the evolution should depend on the
detailed density structure in the outflow-launching region,
which controls the EUV consumption rate.

\section{Simulation Setup}
\label{sec:setup}

As the initial condition of our calculation, we assume the structure of 
a dense core in the run-away collapse from the cosmological simulation 
by {\it (6)}.
The calculation by {\it (6)} followed the entire evolution
from the cosmological initial condition to the birth of a 
primordial protostar under the standard $\Lambda$CDM cosmology.
A small protostar of $M_* \simeq 0.01~M_\odot$ forms 
at $10^{20}~{\rm cm}^{-3}$ as a result of the run-away collapse of a 
dense primordial-gas core at the cosmological redshift $z = 14$. 
Specifically, we take the central 0.3~pc cube around the density peak 
when the maximum density is $10^6~{\rm cm}^{-3}$ as our initial condition. 
We reduce the 3D data to an axisymmetric 2D distribution by averaging 
over azimuthal angles. Our simulation box contains the total gas mass 
of $\simeq 300~M_\odot$. 

The numbers of the grid cells are initially 
$N_Z \times N_R = 42 \times 42$, including 2 ghost cells in each direction.
After we start our axisymmetric calculation, the dense core 
experiences continued gravitational collapse. 
We increase the grid resolution for the central collapsing 
region by successively adding finer nested grids as needed.
With this nested-grid technique, we always resolve the minimum 
Jeans length by tens of grid cells.
The increase of the grid-level is limited up to 8 owing to 
computational cost for following the subsequent accretion phase 
until the final stellar mass is fixed.
We terminate the calculation of the run-away collapse when
the minimum Jeans length becomes too short to be resolved with
the finest grids.
At this point we create a sink cell at the origin
and calculate the subsequent accretion phase as described in Sec.~1.
The central density at this moment is $\simeq 10^{12}~{\rm cm}^{-3}$,
and the cell size at the finest grid-level is $\simeq 12$~AU.

This recipe enables us to smoothly connect the 3D cosmological
simulation using the particle method to a 2D (axial symmetry assumed)
local simulation using the nested-grid method.  
We have confirmed that the evolution in our 2D calculation
is reasonably consistent with the 3D results after the maximum 
density exceeds $10^6~{\rm cm}^{-3}$. 
For example, the upper panel of Figure~\ref{fig:fkep_kprm} shows a
comparison of the angular-momentum profiles for 
the 3D and 2D calculations using
a parameter,
\begin{equation}
\label{eq:fkep}
f_{\rm Kep} (M_r) \equiv \frac{V_{\phi,r}}{V_{\rm Kep,r}}, \qquad
V_{\phi,r} = \frac{l_r}{r}, \
V_{{\rm Kep},r} = \sqrt{ \frac{G M_r}{r} },
\end{equation}
where $M_r$ is the enclosed mass within radius $r$, and 
$l_r$ is specific angular momentum averaged over the spherical shell
whose radius is $r$.
The profile for the 2D simulation is within 20~\% offset from that
in the 3D case taken from {\it (6)}.

Although most of our simulations were done with the settings described 
above, we also calculated early evolution, turning off the stellar feedback,
with the higher maximum grid-level of 9.
The spatial resolution for $R < 240$~AU is doubled in this case and
size of the finest cell is $\simeq 6$~AU.
Figure \ref{fig:xmdot_a} shows the evolution of the accretion rate
until the stellar mass reaches $30~M_\odot$.
The accretion rates are initially a bit lower with the higher central 
resolution but converge to our fiducial case as the stellar
mass increases.

\section{Structure of the Circumstellar Disk}
\label{sec:disk}
 
Here, we examine the detailed structure of the circumstellar 
disk observed in our fiducial case.
Figure \ref{fig:1000au_snapshot} shows the 2D structure of the 
region within 1000~AU of the protostar, when the stellar mass is
$\simeq 10~M_\odot$ and $20~M_\odot$.
The snapshot for the $10~M_\odot$ star shows the structure before
the formation of an HII region.
We see that the accreting gas hits the disk surface and 
heats up to $\simeq 4000$~K by compression. 
Radiative cooling via H$^-$ free-bound emission operates here
as explained in Sec.~\ref{sssec:rad}. 
Figure \ref{fig:disk} displays the radial structure of the disk 
at this moment. 
The hydrogen is primarily in molecular form within the
disk due to rapid formation via the gas phase three-body reaction.
Note that the existence of H$_2$ molecules and their contribution for 
radiative cooling are not included in the semi-analytic models.
The total mass of this molecular gas is $\simeq 10~M_\odot$, which
is comparable to the stellar mass.
Gas on the equator at $R < 1000$~AU has the rotational velocity more 
than 80\% of the local Kepler value. 
The disk scale height is resolved
on average except for the innermost part at $R < 100$~AU.
The profile of the Toomre Q-parameter, a measure of the
gravitational stability of the disk, shows that the minimum
value $Q_{\rm min}$ is almost unity. 
Thus, the disk is marginally gravitationally stable against
the fragmentation. 
However, high-resolution 3D simulations demonstrate that 
the disk fragments due to gravitational instability in
very early stage of the accretion phase {\it (28)}.
Note that ``ring-like'' fragmentation can be captured even in 
axisymmetric 2D simulations.
The absence of fragmentation in our simulation is probably due to 
the limited spatial resolution and the large values
of the viscous $\alpha$-parameter, $\alpha_0 \geq 0.3$ 
(also see Sec.~\ref{sssec:ang}). 
In our case, the disk is also marginally stable in an earlier
stage when the stellar mass is $\simeq 3~M_\odot$ (Fig.~\ref{fig:disk}).
This is almost the same even with the doubled spatial resolution for
$R < 240$~AU.
Even disks which experience fragmentation have a quasi-static structure 
with $Q_{\rm min} \simeq 1$, if averaged over many rotation periods.
Such structure is well mimicked in our calculations.

The right panel in Figure~\ref{fig:1000au_snapshot} shows the
2D disk structure just after the birth of the HII region.
We see a strong photoevaporating flow in the polar directions within
the HII region. The stellar FUV luminosity has also increased  
significantly, but the H$_2$ molecular disk still exists, shadowing the 
stellar photodissociating photons from the dense equatorial regions, 
evident in the disk's temperature and density distributions. The mass 
of the molecular gas is $\simeq 10~M_\odot$ at this moment.

\section{Comparison to Semi-analytic Models}
\label{sec:mt08}
 
Our calculations show that mass accretion toward forming
first stars is shut off via dynamical expansion of the HII region and
photoevaporation of the circumstellar disk. 
This is qualitatively consistent with the picture predicted by 
the semi-analytic models {\it (8)}.
Their models show that the final stellar mass is 
$\simeq 145~M_\odot$ for their fiducial case; this value varies 
with different sets of input parameters.
One of the input parameters is $f_{\rm Kep}$ given by equation
(\ref{eq:fkep}); their adopted fiducial value is $0.5$.
Figure~\ref{fig:fkep_kprm} shows that
$f_{\rm Kep} \simeq 0.6-0.7$ for our fiducial case and is
lower than $0.5$ for the weak-rotation case.
For the semi-analytic model with $f_{\rm Kep} > 0.25$, however, 
the final stellar mass does not depend strongly on $f_{\rm Kep}$, 
but rather on the entropy of the accreting gas,
measured with a non-dimensional parameter
\begin{equation}
K' \equiv \frac{P / \rho^\gamma}{1.88 \times 10^{12} {\rm cgs}}
= \left( \frac{T}{300~{\rm K}} \right)
  \left( \frac{10^4 {\rm cm}^{-3}}{n} \right)^{0.1} ,
\end{equation}
where $\gamma = 1.1$ is adopted for a typical value for
the primordial gas {\it (2)}.

The semi-analytic models use $K'=1$ for their fiducial case and
show that the final mass roughly scales as $K'^{1.3}$.
From Figure \ref{fig:fkep_kprm} we see that
$K' \simeq 0.7$ for our fiducial case and $K' \simeq 0.8 - 1$ for
the weak-rotation case. 
The semi-analytic model with $K' = 0.7$ predicts a final mass 
$\simeq 90~M_\odot$, twice our value of $45~M_\odot$.
The final mass $\simeq 85~M_\odot$ for our weak-rotation case is
somewhat lower than the semi-analytic prediction $> 108~M_\odot$
for $K' > 0.8$.

The initial conditions in our examined cases correspond 
to typical values for gas clouds bearing the first stars.
Figure~\ref{fig:xmdot_a} shows that the accretion rates
for the cases considered lie in the range
$10^{-2}$ to $10^{-3}~M_\odot~{\rm yr}^{-1}$ for the 
$10~M_\odot$ star, comparable to that 
expected for 12 mini-halos using the semi-analytic models \cite{TSO10}.
However, our calculations attain systematically lower final
masses than predicted by semi-analytic models.

We examine possible reasons for this difference below. 
First, in our calculations, the mass ratio between the star and disk
$f_d \equiv M_{\rm disk}/M_*$ differs from the fiducial value in the 
semi-analytic models, $f_d = 0.3$.
As discussed in Sec.~\ref{sec:disk}, for our fiducial case the
mass of the molecular disk is $\simeq 10~M_\odot$, when the 
stellar mass is $10~M_\odot$, and would be even higher, if we 
include the outer atomic part.
In 2D simulations, however, the value of $f_d$ depends on 
the degree of $\alpha$-viscosity, $\alpha_0$ (see Sec.~\ref{sssec:ang}).
With the larger value $\alpha_0 = 1$, for example, the mass of the 
molecular disk for the $10~M_\odot$ star is reduced to $8~M_\odot$.
To derive the appropriate value of $f_d$, we need detailed
3D numerical simulations which solve the transport of angular
momentum in self-gravitating disks.
Recent work on the Galactic star formation demonstrates that
massive disks with $f_d \geq 1$ can form in some situations, though 
the value of $f_d$ varies with different initial conditions 
of gas cores \cite{Mcd10,Kratter10}.

We contend, however, that differences in the stellar feedback process 
are principally responsible for the differences in final stellar mass. 
The semi-analytic models adopt the following formula for estimating
the photoevaporation rate of the disk,
\begin{equation}
\dot{M}_{\rm evp} = 4.1 \times 10^{-5}
~\left( \frac{S_{\rm EUV}}{10^{49}~{\rm sec}^{-1}} \right)^{1/2} 
\left( \frac{T_{\rm HII}}{10^4~{\rm K}} \right)^{2/5} 
\left( \frac{M_{\rm *d}}{100~M_\odot} \right)^{1/2}
\ \ M_\odot~{\rm yr}^{-1} ,
\label{eq:xmdot_evp}
\end{equation}
where $M_{* d} = M_* + M_{\rm disk}$ \cite{HJLS94}.
For our fiducial case the photoevaporation rate is about
$2 \sim 3\times 10^{-4}~M_\odot~{\rm yr}^{-1}$, after the stellar mass
exceeds $30~M_\odot$.
By contrast, equation (\ref{eq:xmdot_evp}) estimates an evaporation
rate of several $10^{-5}~M_\odot~{\rm yr}^{-1}$,
using $T_{\rm HII} = 5 \times 10^4$~K and the values of $S_{\rm EUV}$
and $M_{\rm *d}$ from the simulation ($M_{\rm disk}$
is the mass of the molecular disk).
We attribute this difference to the basic assumption of an infinitely 
thin disk for
deriving equation (\ref{eq:xmdot_evp}) \cite{HJLS94}.
As a result, only the diffuse EUV radiation (see Sec.~\ref{sssec:rad})
is accounted for in equation (\ref{eq:xmdot_evp}). 
In our calculations the stellar direct EUV radiation field impinging on
the flared disk is primarily responsible for determining
the photoevaporation rate.
Even if the transfer of diffuse EUV radiation is turned off
in the simulation, we do not see any significant change in
the evolution.

Moreover, the semi-analytic models assume that,
after the HII region forms,
mass accretion onto the circumstellar disk still continues
from regions shaded by the disk, where the stellar UV radiation is blocked.
In this picture the stellar final mass is determined when
the photoevaporation rate exceeds the mass
accretion rate from the shaded regions.
This does not agree with our simulations.
It is true that the stellar UV radiation is blocked behind
the disk as shown in Figure 2 in the main article.
In our calculations, however, a shock front propagates into the
accretion envelope shaded by the disk, as the HII region 
dynamically expands.
Figure~\ref{fig:shots_pn} shows that there is the outward pressure gradient 
within the HII region due to the photoevaporating flow.
As the HII region dynamically expands, the shocked 
accretion envelope behind the disk also obtains the same 
pressure gradient.
As a result, the shocked gas is accelerated outward due to this
pressure gradient at a velocity of several km/sec.
This is shown in Figure \ref{fig:vrz0}. We see that the gas
at $R >$ a few $10^3$~AU is moving outward when the
stellar mass exceeds $20~M_\odot$.
The outflow rate of atomic gas via this horizontal motion is
$9 \times 10^{-4}~M_\odot~{\rm yr}^{-1}$, when the stellar mass
is $35~M_\odot$.
The mass supply from the envelope to the disk is cut off by this
moment.\footnote{In our simulations, we adopt a 
semi-permeable outer boundary condition, which allows outflow 
but prohibits mass inflow through the boundary.
In principle this boundary condition can ultimately limit mass 
accretion onto the disk. However, mass infall toward the disk is
reversed and the gas flows outward through the outer boundary
long before this limit is reached.}
The isolated disk without replenishment from the envelope continues
to lose gas by the photoevaporation. Ultimately, mass accretion
onto the protostar ceases and the stellar final mass
is fixed.

\section{Potential Effects of 3-Dimensionality}

For this study we have adopted an axisymmetric 2D code for
calculating the dynamical evolution of the accreting flow.  
Recent 3D simulations for the present-day high-mass star
formation are helpful for speculating on potential 3D effects.

A primary radiative feedback effect from Galactic high-mass stars
is radiation pressure exerted on dust grains coupled with the gas 
accretion flow. 
Historically, this effect was first studied under spherical symmetry 
\cite{Kahn74,YK77,WC87} and it was
found that the accretion flow toward high-mass stars is prevented 
by the strong repulsive force (called as "radiation pressure problem").
2D numerical simulations show that non-spherical mass accretion 
via disks significantly alleviates this problem \cite{YS02,kuiper10}.
3D simulations by \cite{Krum09} demonstrate that 3D effects
(e.g., radiative Rayleigh-Taylor instability growing in the 
accretion envelope) also reduce the repulsive force in addition
to the effect of the disk accretion.
However, it is still controversial whether the 3D effects are
essential for determining the upper mass limit for 
present-day high-mass star formation.
More recent 3D simulations by \cite{kuiper11} show that the radiation 
pressure barrier is circumvented with the disk accretion, but find no 
critical 3D effects. 

Without dust grains the opacity of the primordial gas is much lower
than for the present-day Galactic interstellar medium.
The dynamical evolution discussed in our article is 
not primarily due to radiation pressure, but rather to the
high gas pressure of photoionized gas, which causes the dynamical
expansion of the HII region supplemented by the photoevaporation 
of the disk.
UV feedback effects in the present-day high-mass star forming regions 
have also been studied with 3D numerical simulations \cite{Peters10a, Dale05}.
However, these studies focus on the feedback effects over the large
length-scale of cluster-forming clumps, $\sim 1$~pc. 
By contrast, we consider here the smaller-scale feedback processes in the
vicinity of protostars, such as photoevaporation of the disks.
The photoevaporation rate from the disk is expected to be highest at a radial
distance from the star $R = R_{g,{\rm HII}}$ given by equation
(\ref{eq:rghii}) \cite{HJLS94}.
However, the regions smaller than $1000$~AU around forming stars are 
not spatially resolved in the above-cited studies.
The photoevaporation of the disk is a local process which
takes place in the very central part of the HII region.
In 3D we can expect the photoionized gas 
to have a clumpier structure than modeled by our 2D simulations. 
However, such small-scale inhomogeneities would be smoothed out 
within the sound crossing time, which is much shorter than the dynamical 
timescale of the expanding HII region. 

As referred in Sec.~\ref{sssec:ang}, recent 3D simulations
demonstrate that the circumstellar disks around primordial 
protostars fragment in the early stage of the accretion phase.
The fragmentation of the disk would somewhat reduce the stellar final 
masses, as the accreting materials are shared among the fragments
\cite{Peters10b}. 
Several authors have also suggested that a turbulent velocity field is 
present after the gravitational collapse of primordial gas cores {\it (3)}. 
Although the expected turbulent field is not strong for the very
first stars, we note that, in our case, the turbulent field is further
weaken by mapping the 3D data to the 2D axisymmetric data
(see Sec.~\ref{sec:setup}). 
With a random velocity field the angular momentum directions 
of accreting materials would be time-dependent. As a result, the 
polar direction of the circumstellar disk should vary with time as well
and the photoevaporating flow would clear out a larger amount
of accreting materials than otherwise.
We speculate that this effect would accelerate the horizontal 
expansion of the HII region and reduce the resulting final stellar mass.
Our obtained final mass of the first stars,
several $10~M_\odot$, is thus a conservative upper limit.
Nonetheless, this is still much lower than the postulated high
values exceeding $100~M_\odot$.

\clearpage
\setcounter{figure}{0}
\renewcommand{\thefigure}{S\arabic{figure}}

\begin{figure}
\begin{center}
\psfig{file=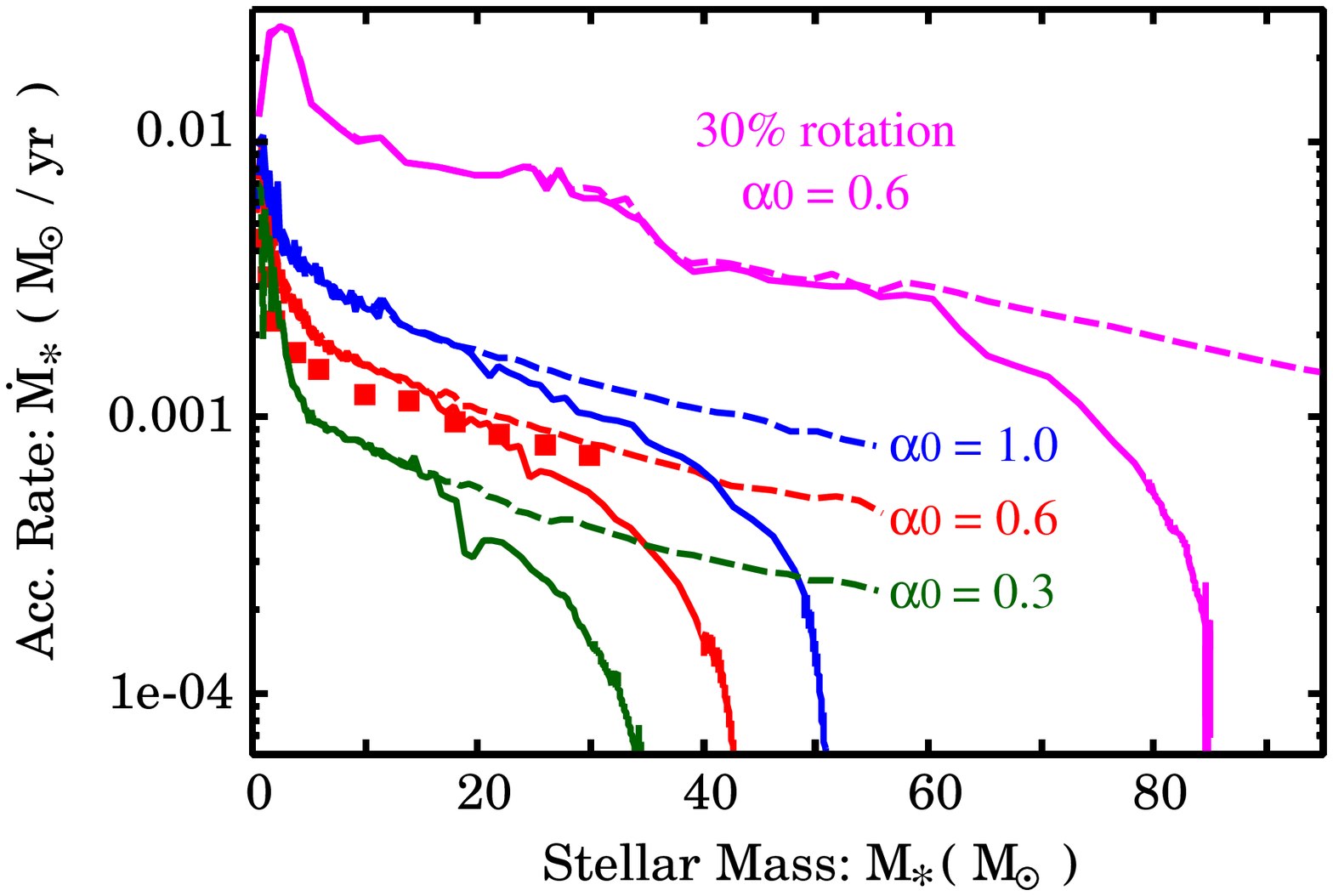,width=0.9\textwidth}
\caption{Evolution of accretion rates onto the protostar
with different $\alpha$-parameters for angular momentum
transport. The blue, red, and green curves depict the 
results with $\alpha_0 = 1$, 0.6, and 0.3 in equation
(\ref{eq:alpha}), respectively.
The magenta curves display the evolution with the initial
angular momentum reduced to 30\% of the fiducial value.
For each case the solid and dashed lines represent the evolution with 
and without radiative feedback from the protostar, 
respectively.
The red filled squares represent the no-feedback case with 
$\alpha_0 = 0.6$ doubling the spatial resolution in the central 
region of $R < 240$~AU (also see Sec.~\ref{sec:setup}).
}
\label{fig:xmdot_a}
\end{center}
\end{figure}
\begin{figure}
\begin{center}
\psfig{file=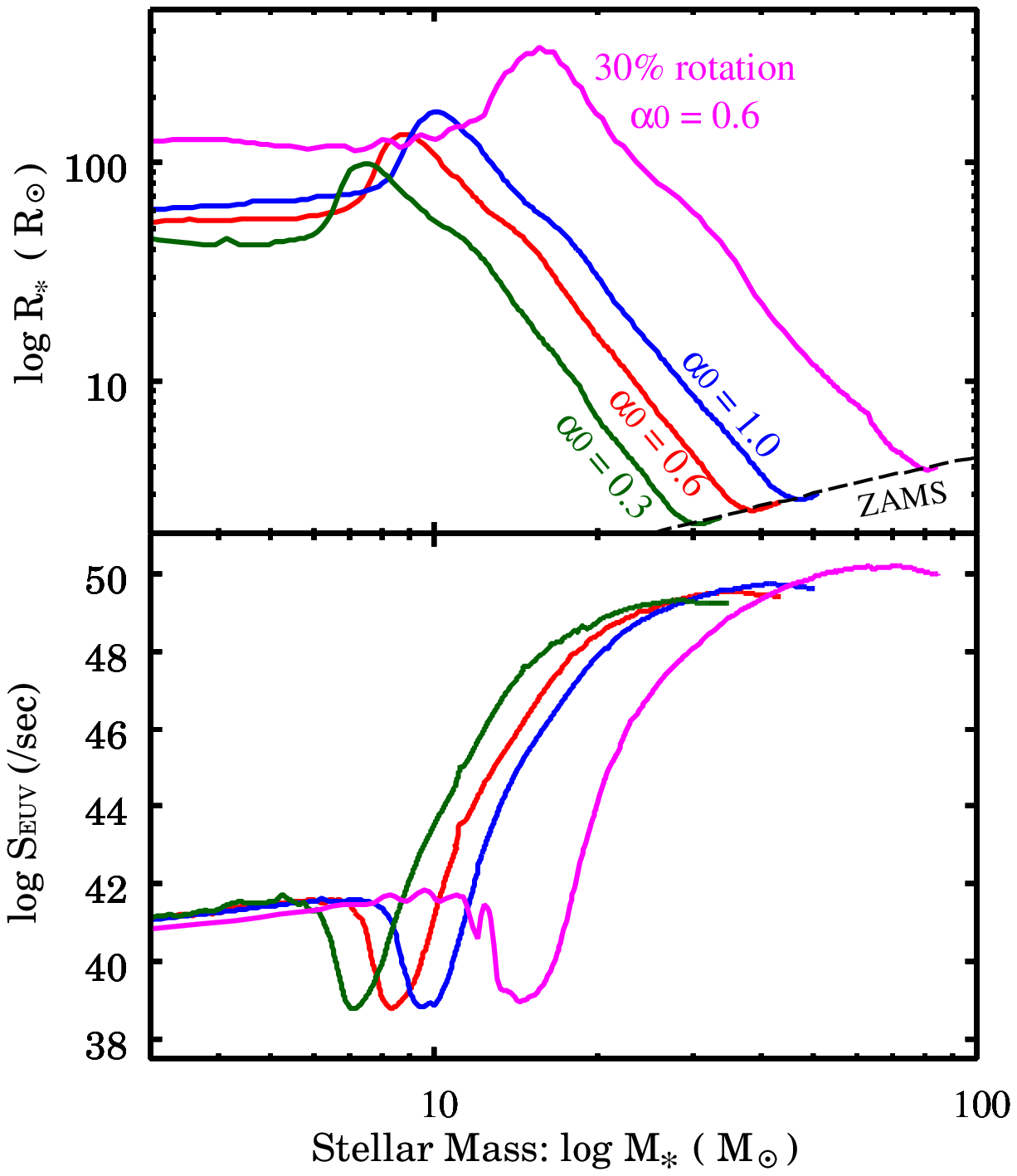,width=0.9\textwidth}
\caption{Accretion rates onto the protostar as a function of
the accreted stellar mass, assuming
different $\alpha$-parameters for angular momentum
transport. The blue, red, and green curves depict the evolution
with $\alpha_0 = 1$, 0.6, and 0.3 in equation
(\ref{eq:alpha}), respectively.
The magenta curves display the evolution with the initial
angular momentum reduced to 30\% of the fiducial value.
For each case the solid and dashed lines represent the evolution with 
and without radiative feedback from the protostar, 
respectively.}
\label{fig:pevol_a}
\end{center}
\end{figure}
\begin{figure}
\begin{center}
\psfig{file=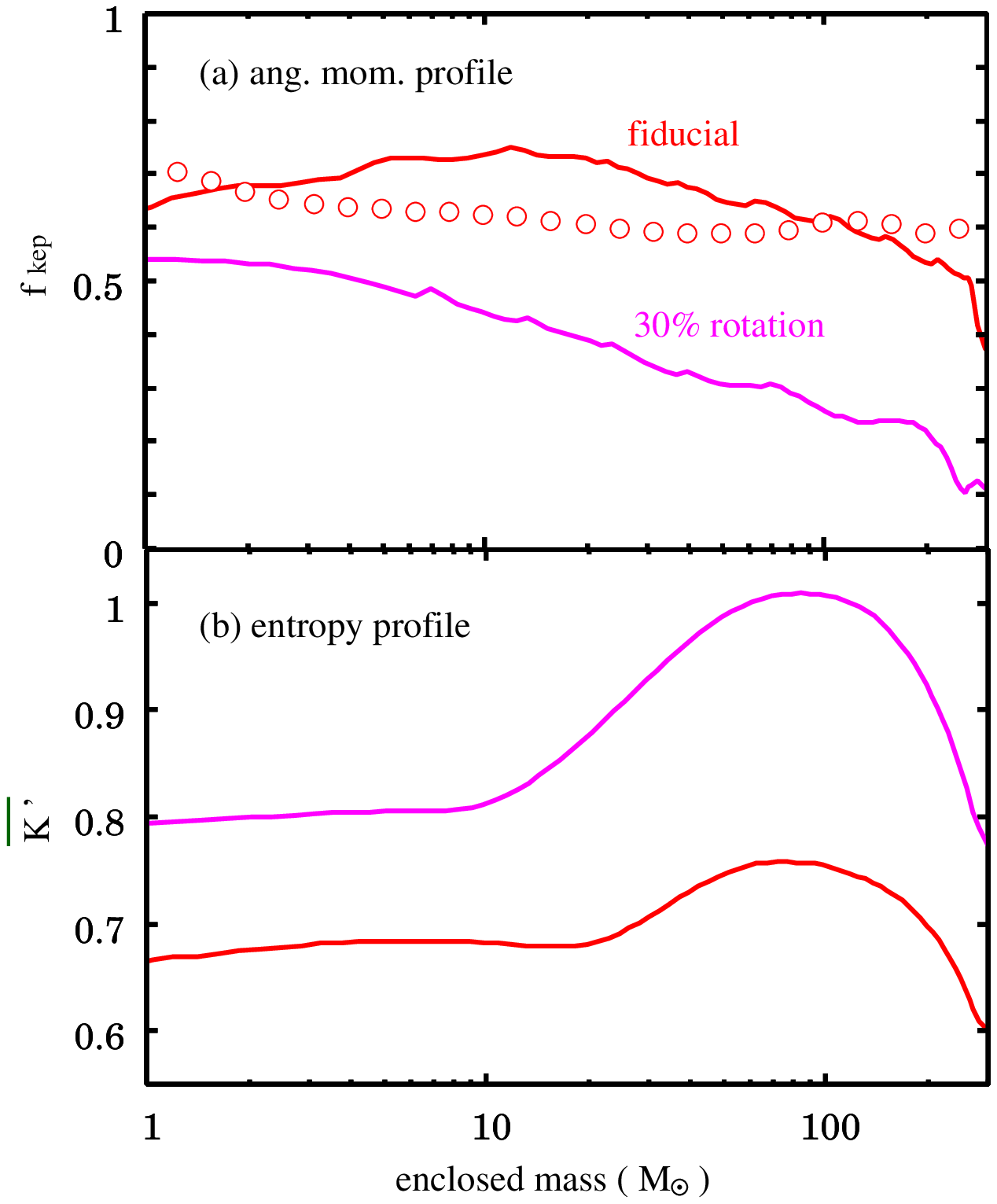,width=0.9\textwidth}
\caption{Structure of the accreting envelope
at the end of the calculation of the run-away collapse stage for
the fiducial case (red lines) and weak-rotation case (magenta lines).
{\it Upper panel:} the ratio of the local rotational velocity
to Kepler velocity $f_{\rm kep}$.
The red open circles represent a snapshot at the birth of a embryo
protostar in a three-dimensional cosmological simulation {\it (6)}.
{\it Lower panel:} Dimensionless entropy of the
accreting gas $K'$. The mass-weighted average values $\overline{K'}$
are plotted against the enclosed gas mass.}
\label{fig:fkep_kprm}
\end{center}
\end{figure}
\begin{figure}
\begin{center}
\psfig{file=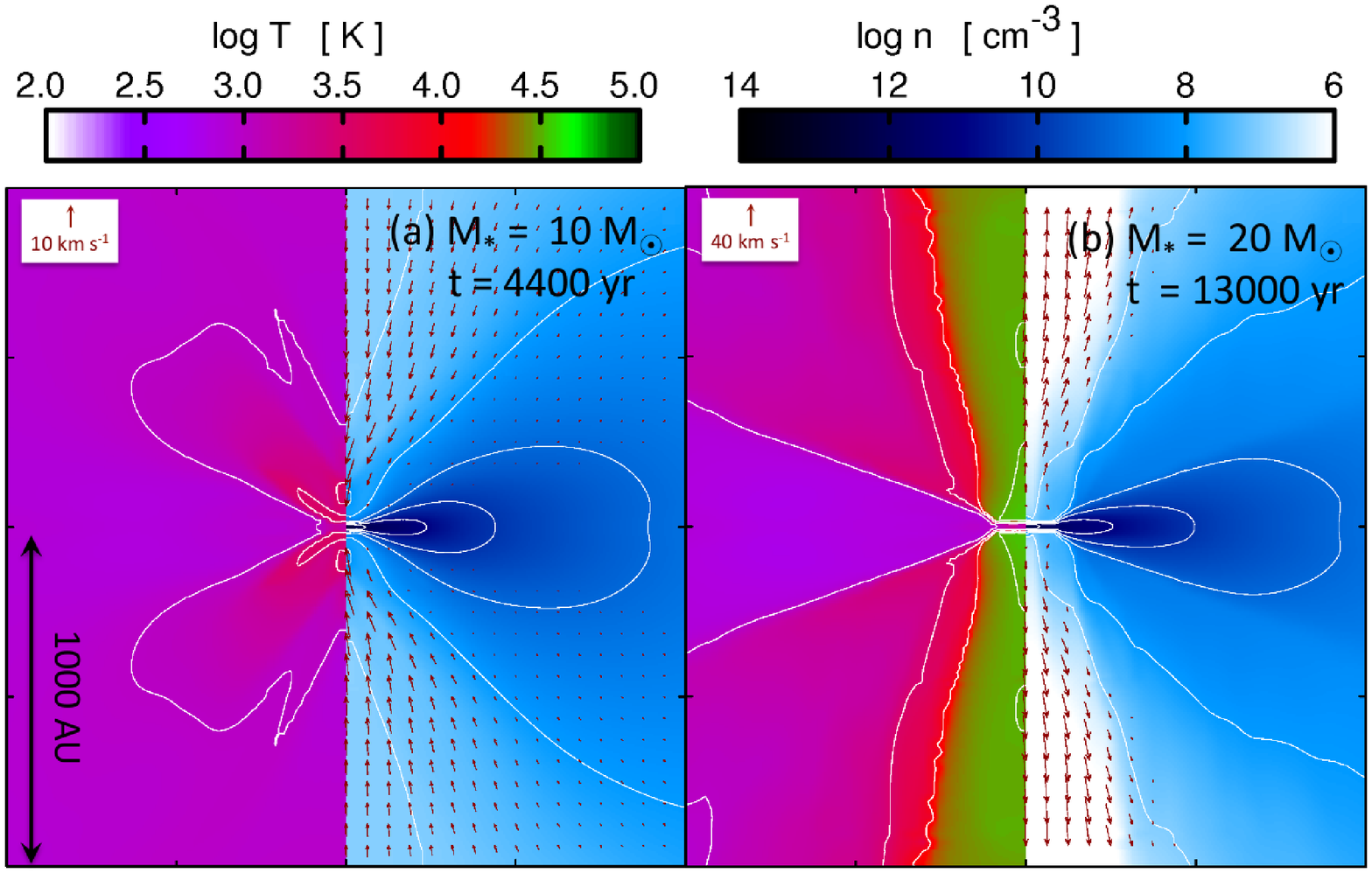,width=1.0\textwidth}
\caption{
The structure of gas temperature (left-hand side) and of number 
density and velocity (right-hand side) in the vicinity of $1000$~AU 
around the protostar.
The left and right panels show the snapshots when the stellar
mass is  $\simeq 10~M_\odot$ and $20~M_\odot$.
The elapsed time since the birth of the protostar is also shown
in each panel.
Note that the color scale of the density, the legend of the velocity
vectors, and size of the plotted area are different from those in
Figure 2 in the main article.
}
\label{fig:1000au_snapshot}
\end{center}
\end{figure}
\begin{figure}
\begin{center}
\psfig{file=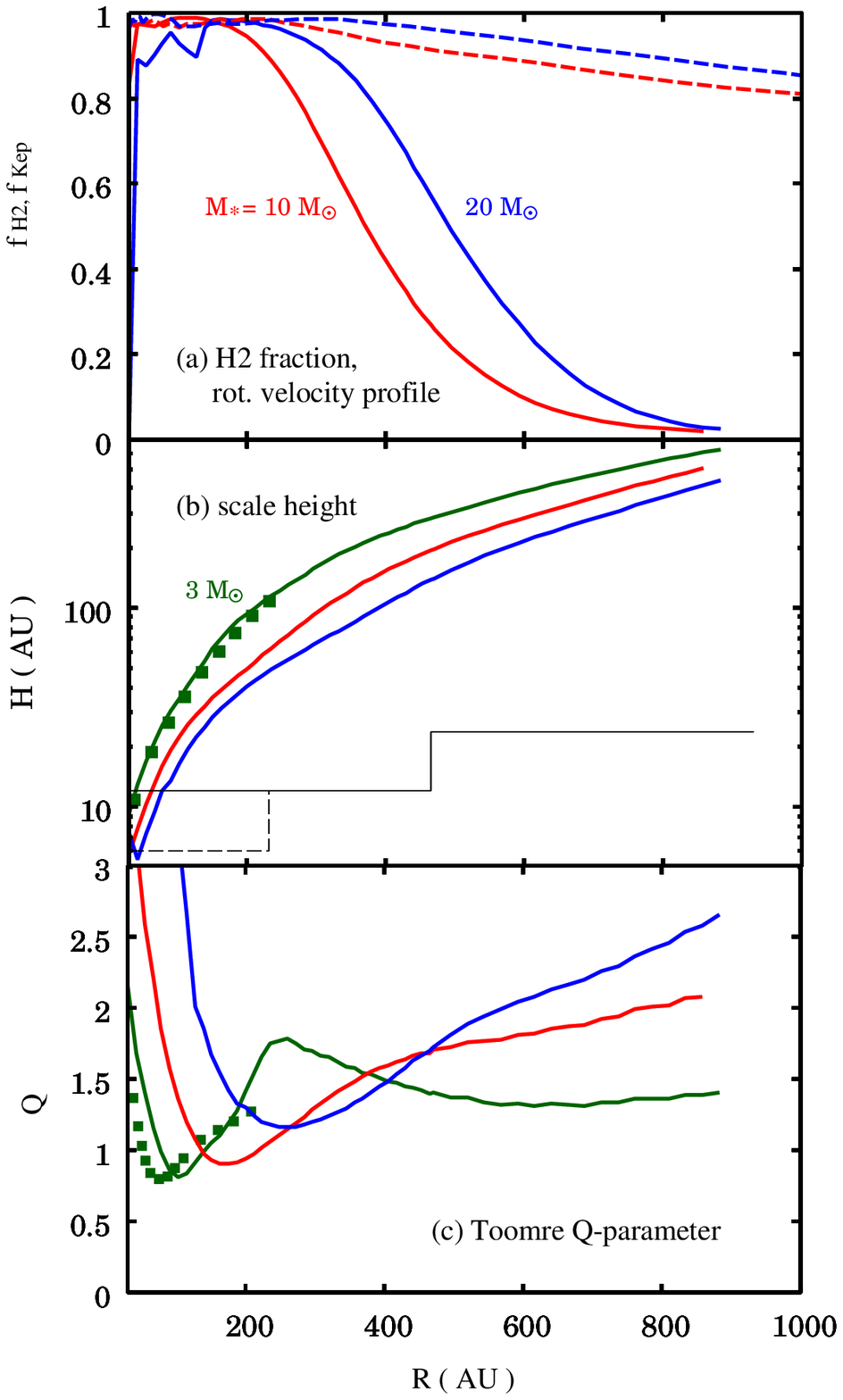,width=0.6\textwidth}
\caption{Radial structure of the circumstellar disk 
within $1000$~AU around the protostar when 
the stellar mass is $3~M_\odot$ (green lines for the lower two panels), 
$10~M_\odot$ (red lines) and $20~M_\odot$ (blue lines). 
{\it Upper panel:} The fraction of hydrogen molecules along
the equator $f_{\rm H_2}$ (solid lines), and the ratio 
of the local rotational velocity to Kepler velocity 
$f_{\rm Kep}$ (dashed lines).
{\it Middle panel:} The scale height of the disk $H$. The black
thin line shows the grid resolution in our standard cases, and the 
discontinuity at $R \simeq 470$~AU indicates a grid-level boundary there.
The profile when the stellar mass is $3~M_\odot$
with the doubled spatial resolution for $R < 240$~AU
is also presented (green filled squares). 
The black thin dashed line shows the grid resolution in this case.
{\it Lower panel:} Same as the middle panel but for 
Toomre Q-parameter at each radius.
}
\label{fig:disk}
\end{center}
\end{figure}
\begin{figure}
  \begin{center}
\psfig{file=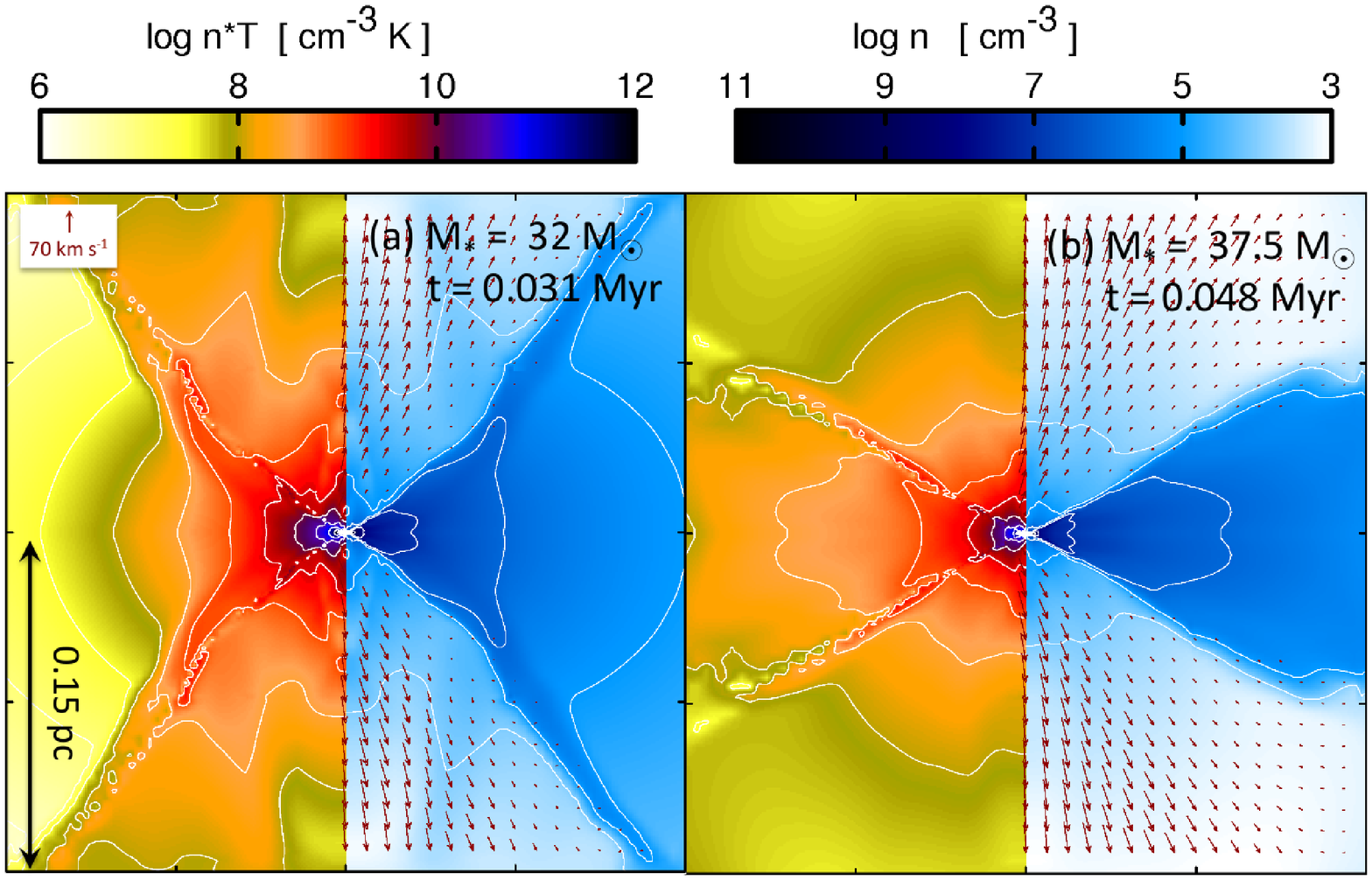,width=1.0\textwidth}
\caption{Same as Fig.~2 in the main article but for gas temperature
multiplied by number density in the left panel, which is 
nearly proportional to gas pressure.
The left and right panels show snapshots at times, when the stellar 
mass is $M_* = 32~M_\odot$ and $37.5~M_\odot$ respectively.
}
\label{fig:shots_pn}
  \end{center}
\end{figure}
\begin{figure}
\begin{center}
\psfig{file=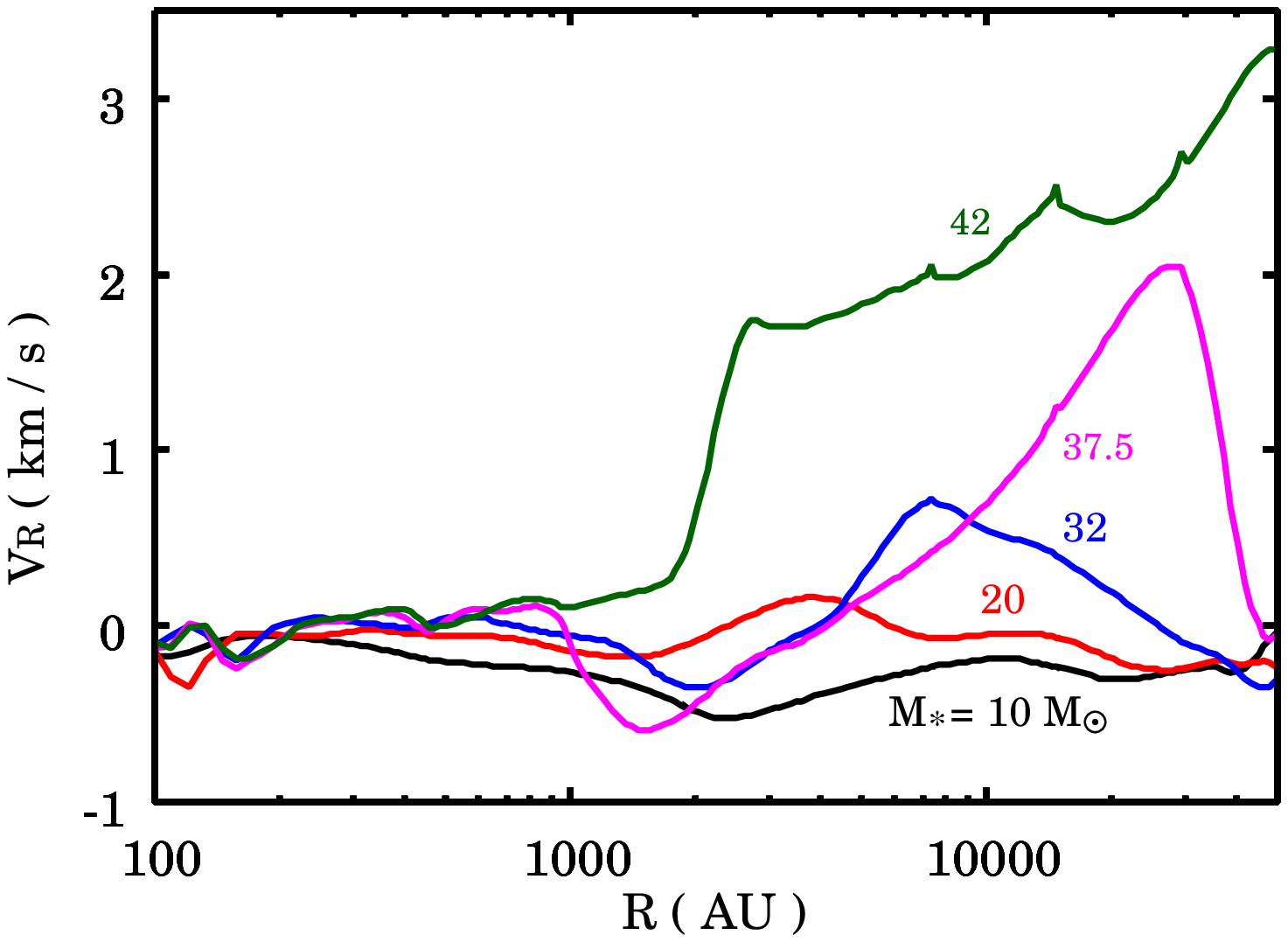,width=1.0\textwidth}
\caption{Distribution of $R$-component of the radial velocity $v_R$
along the equator ($Z=0$).
The snapshots when the stellar mass is $10~M_\odot$ (black), 
$20~M_\odot$ (red), $32~M_\odot$ (blue), $37.5~M_\odot$ (magenta), 
and $42~M_\odot$ (green) are plotted.
Positive $v_R$ means that gas is moving outward. 
}
\label{fig:vrz0}
\end{center}
\end{figure}

\end{document}